\documentclass[11pt,a4paper]{article}
\usepackage{jinstpub}
\usepackage{lineno}

\graphicspath{{./Figures/}}
\usepackage[utf8]{inputenc}
\usepackage{lipsum}

\usepackage[T1]{fontenc}
\usepackage{mathtools}
\usepackage{enumitem}
\usepackage{overpic}
\usepackage{subfig}
\usepackage{wrapfig}
\usepackage{placeins}
\usepackage[export]{adjustbox}
\usepackage{tabularx}

\usepackage{upgreek}
\usepackage{flafter}
\usepackage{flushend}

\usepackage[dvipsnames]{xcolor}
\definecolor{rltred}{rgb}{0.75,0,0}
\definecolor{rltgreen}{rgb}{0,0.5,0}
\definecolor{rltblue}{rgb}{0,0,0.5}

\usepackage{bm}
\usepackage{rotating}
\usepackage{caption}
\usepackage{hhline}
\usepackage{threeparttable}
\usepackage{array,multirow,dcolumn}
\newcommand\mc[3]{\multicolumn{#1}{#2}{#3}}
\newcommand{\br}[1]{$\langle#1\rangle$}
\newcolumntype{=}[1]{D{=}{\,=\,}{#1}}

\newcommand{\SSD}{\href{https://github.com/JuliaPhysics/SolidStateDetectors.jl}{\mbox{\textit{SolidStateDetectors.jl}}}}
\newcommand{\lsq}{\href{https://github.com/JuliaNLSolvers/LsqFit.jl}{\mbox{\textit{LsqFit.jl}}}}
\newcommand{\RNum}[1]{\uppercase\expandafter{\romannumeral #1\relax}}
\newcommand{\tbf}{\textbf}

\title{Temperature Dependence of the Electron-Drift Anisotropy and Implications for the Electron-Drift Model}
\author[a]{I.~Abt}
\author[a]{C.~Gooch}
\author[a]{F.~Hagemann}
\author[a]{L.~Hauertmann}
\author[b,c,a]{D.~Hervas Aguilar}
\author[a]{X.~Liu}
\author[a]{O.~Schulz}
\author[a,1]{M.~Schuster \note{Corresponding author}}
\author[a]{A.J.~Zsigmond}

\affiliation[a]{Max-Planck-Institut für Physik,\\
            F\"ohringer Ring 6, Munich,
            80805, Germany}
            
\affiliation[b]{University of North Carolina, Department of Physics and Astronomy,\\
            120 E. Cameron Ave., Phillips Hall CB3255, 
            Chapel Hill,
            27599,
            NC,
            USA}
            
\affiliation[c]{Triangle Universities Nuclear Laboratory,\\
            116 Science Drive, Duke University, 
            Durham,
            27708,
            NC,
            USA}

\emailAdd{schuster@mpp.mpg.de}

\abstract{The electron drift in germanium detectors is modeled making many assumptions. Confronted with data, these assumptions have to be revisited. The temperature dependence of the drift of electrons was studied in detail for an n-type segmented point-contact germanium detector.
 The detector was mounted in a temperature controlled, electrically cooled cryostat. Surface events were induced with collimated 81\,keV photons from a $^{133}$Ba source. A detailed analysis of the rise time of pulses collected in surface scans, performed at different temperatures, is presented. The longitudinal anisotropy of the electron drift decreases with rising temperature. A new approach, making use of designated rise-time windows determined by simulations using $SolidStateDetectors.jl$, was used to isolate the longitudinal drift of electrons along different axes to quantify this observation. The measured temperature dependence of the longitudinal drift velocities combined with the standard electron drift model as widely used in relevant simulation packages results in unphysical predictions. A first suggestion to modify the electron-drift model is motivated and described. The results of a first implementation of the modified model in $SolidStateDetectors.jl$ are shown. They describe the data reasonably well. A general review of the model and the standard input values for mobilities is suggested.}
 
 \keywords{Charge carrier drift, Electron mobility, Experimental setup, High-purity germanium detectors, Pulse shape library, Pulse shape simulation, Rise-time analysis, Temperature Dependence }
 
\begin{document}
\maketitle
\flushbottom

\section{Introduction} \label{sec:introduction}
Germanium detectors are used in a number of different areas in industry and research. In many of these applications, e.g.~the searches for neutrinoless double-beta decay\,\cite{Agostini2020:FinalResultsGERDA,Majorana2019:Searchneutrinolessdouble,LEGEND2017:TheLEGENDExperiment,LEGEND2021:PCDR} or dark matter\,\cite{CoGeNT2013:dmsearch,SuperCDMS2014:wimpsearch,CDEX2018:wimplimits}, not only energy spectra are of interest but also the shape of the charge pulses obtained for the individual energy depositions forming an event. The pulse shape registered on a given detector electrode for an event is determined by the trajectories of the charge carriers, i.e.~the electrons and holes, drifting within the crystal. The charge drift, in turn, is largely governed by the mobilities of the charge carriers, which depend on properties like the crystal structure and the electron density distribution in momentum space. The mobilities are also temperature dependent.\\
Knowledge on the individual temperature dependence of both the electron and hole mobility along the major crystallographic axes and on all the related effects and properties would be extremely valuable to better understand the charge carrier drift in germanium detectors. A thorough understanding of these matters is important for the analysis of pulse shapes. This knowledge can also be used to improve and tune pulse-shape simulations, which are essential to many basic physics analyses.
A study of the temperature dependence of the electron mobility for different crystallographic axes is presented together with a first attempt to describe the results by simulation.

\section{Charge Carrier Drift in Germanium Detectors}\label{sec:ccdrift_ge}
Germanium is one of many different semiconductor materials out of which solid-state detectors can be fabricated. The general working principle is the same for all of them\,\cite{Knoll2010}. Their individual properties and characteristics, however, depend on the respective material. Appendix~\ref{app:germanium_properties} provides a summary of the basic material properties which are important for the understanding of germanium detectors.\\
Apart from the mobilities, the charge carrier drift depends on the static electric field in the detector and the placement of the electrodes. The electric field is the result of a combination of the charge density caused by ionized impurities and the applied reverse-bias voltage, $U_{RB}$. When the electric-field lines are aligned with a crystallographic axis, the charge carriers can drift along the electric-field lines and the longitudinal velocity of either electrons or holes, $v_{e/h}$, is 
\begin{linenomath}
\begin{equation}\label{eq:drift_velocity_simple}
    v_{e/h} = \mu_{e/h} \cdot \mathcal{E}~~~~~,
\end{equation}
\end{linenomath}
where $\mathcal{E}$ is the electric-field strength and $\mu_{e/h}$ is the electron/hole mobility.\\

Typical field strengths in germanium detectors are of the order of $\mathcal{E}_{typ} \approx 1000$\,V/cm\,\cite{Knoll2010}. Equation~\eqref{eq:drift_velocity_simple} only holds for $\mathcal{E} \leq 100$\,V/cm. When the electric field gets stronger, the velocity does not increase linearly any longer but approaches a saturation velocity. This effect can be taken into account by introducing additional model parameters\,\cite{Trofimenkoff1965Fielddependentmobility,Caughey1967:Mobilities,Canali1975:DriftVelocity,Mihailescu:2000jg}:
\begin{linenomath}
\begin{equation}\label{eq:drift_velocity}
    v_{e/h} = \frac{\mu_{e/h}\mathcal{E}}{(1 + (\frac{\mathcal{E}}{\mathcal{E}_{0, e/h}})^{\beta_{e/h}})^{1/\beta_{e/h}}} - \mu_n \mathcal{E}~~~~~,
\end{equation}
\end{linenomath}
where $\beta$ and $\mathcal{E}_{0}$ are temperature-dependent parameters which can be obtained from measurements along the crystal axes. The negative term containing $\mu_n$ accounts for the so-called Gunn effect\,\cite{Ottaviani1975:cct}: At high electric-field strengths, electrons can get excited into higher-energy states and acquire different effective masses. The term containing $\mu_{n}$ amounts to a $\approx 5\%$ correction at $\mathcal{E}_{typ}$.\\  
Table~\ref{tab:mobilities} lists values\,\cite{Bruyneel2006:CharacterizationII} for mobilities, $\mu_{e/h}$, and the model parameters, $\mathcal{E}_{0, e/h}, \beta_{e/h}, \mu_{n}$, as used in simulations performed for this publication, for the \br{100} and \br{111} axes at a temperature of $T = 78$\,K. The values of $\mu_{e/h}$ for these two axes differ only slightly for both electrons and holes. This results in a quasi-isotropic behavior of electrons and holes for low electric-field strengths. The values for the model parameter $\beta_{e/h}$, however, differ significantly for the two axes. Thus, for $E_{typ}$, the drift velocity also significantly differs for the different axes. 
\begin{table}[htb]
    \centering
    \caption[Mobilities and model parameters for charge carrier drift along the $\langle100\rangle$ and $\langle111\rangle$ axes.]{Mobilities and model parameters as defined in Eq.\,\eqref{eq:drift_velocity} for charge carrier drift along the $\langle100\rangle$ and $\langle111\rangle$ axes. The parameters are results from fits to experimental data at $T = 78$\,K\,\cite{Bruyneel2006:CharacterizationII}.}
    \begin{tabular}{lcccr}
        &\multicolumn{2}{c}{\tbf{Electrons}} & \multicolumn{2}{c}{\tbf{Holes}}\\
        \tbf{Parameters}&$\langle100\rangle$ & $\langle111\rangle$ & $\langle100\rangle$ & $\langle111\rangle$\\
        \hhline{=====}
        $\mu_{e/h}$ in cm$^2$/Vs & 38609 & 38536 & 61824 & 61215\\
        $\mathcal{E}_{0, e/h}$ in V/cm & 511 & 538 & 185 & 182\\
        $\beta_{e/h}$ & 0.805 & 0.641 & 0.942 & 0.662\\
        $\mu_n$ in cm$^2$/Vs & -171 & 510 & 0 & 0\\
        \hline
    \end{tabular}
    \label{tab:mobilities}
\end{table}\\
As measured parameters for Eq.\,\eqref{eq:drift_velocity} are only available for two crystallographic axes, a drift model is needed to make any prediction for the drift on the third axis, $\langle110\rangle$, or between the axes. The electron drift velocity for the studies presented here is generally calculated following the approach of Mihailescu~et~al.\,\cite{Mihailescu:2000jg}. Appendix\,\ref{app:e_drift_model} provides the concepts and results of this model. In germanium, the relation $v_{e}^{100} \geq v_{e}^{110} \geq v_{e}^{111}$ is observed in general.\\

\label{sec:tdep}%
The mobilities depend on the different scattering processes in the crystal, which individually depend on the respective effective masses for electrons and holes\,\cite{Nag1980:ElectronTransport}, see App.\,\ref{app:germanium_properties}. In other words, the influence of each scattering process, $s$, represents an effective increase in resistivity.  The $ansatz$ is to absorb all these effects into $\mu_{e/h}$. The respective contributions, $\mu^{s}_{e/h}$, are summed up via Matthiessen's rule\,\cite{Anderson1986:mobilitySum}:
\begin{linenomath}
\begin{equation}\label{eq:matth}
    \frac{1}{\mu_{e/h}} = \sum\limits_{s} \frac{1}{\mu^s_{e/h}}~~~~~.
\end{equation}
\end{linenomath}
 The overall value of $\mu_{e/h}$ and its temperature dependence\,\cite{Prince1953:DriftMobilitiesSemiconductors}, hence, is given by the superposition of these scattering processes, as each $\mu^{s}_{e/h}$ has a characteristic dependency on $T$ and $m^{*}_{e/h}$. In terms of effective masses, higher $m^{*}_{e/h}$, in most cases, result in lower $\mu^{s}_{e/h}$.
For electrons, the following scattering centers have to be considered:
\begin{itemize}
    \item \textbf{Acoustic phonons, $aco$:} When an electron transfers part of its momentum to a nucleus in the crystal lattice, the nucleus starts to oscillate in the Coulomb potential of the surrounding atoms. As this oscillation propagates to the neighboring nuclei through Coulomb interactions, it can be described as a wave. The corresponding quasi-particles are called phonons. When the base atoms of a primitive cell move coherently, the resulting phonons are called acoustic phonons. Acoustic phonons can occur in any crystal. The mobility contribution due to acoustic phonons, $\mu^{aco}_{e}$, is calculated to be proportional to \mbox{$(m^*_{e})^{-5/2}\cdot T^{-3/2}$}\,\cite{Bardeen1950:DeformationPotentialsMobilities}.
    \item \textbf{Optical phonons, $opt$:} In crystals with bases consisting of two or more atoms, optical phonons can occur. The atoms of the base move out-of-phase. Germanium has a two-atomic base with covalent bonds between the atoms and optical phonons do occur. The contribution to the mobility from interactions with optical phonons is also temperature dependent. However, it is negligible for typical germanium detector operating temperatures as it is only relevant for $T > 300$\,K\,\cite{Mei2016:impactneutralimpurity}.
    \item \textbf{Ionized impurities, $ion$:} Ionized impurities are electrically active. They mainly consist of the dopants included in the crystal and their density is typically of the order of $\rho \approx 10^{9}-10^{10}$\,cm$^{-3}$ in germanium detectors.
    The mobility contribution from ionized impurities, $\mu_{e}^{ion}$, scales with $\frac{(m^*_{e})^{-1/2}}{\rho} \cdot T^{3/2}$\,\cite{Conwell1950:ImpurityScattering}. 
    \item \textbf{Neutral impurities, $neu$:} In addition to the ionized impurities, also neutral impurities are found in the crystals. There are many sources of neutral impurities.
    Traces of e.g.~thorium, uranium, copper or gold are already contained in the germanium ore. During the crystal purification procedure, carbon can be introduced from a graphite boat. During the crystal pulling, hydrogen, which is used as gas inside the puller, as well as silicon and oxygen from the quartz-based crucible diffuse into the crystal. Typical concentrations of neutral impurities, $\rho_{neu}$, are of the order of $10^{14}-10^{15}$\,cm$^{-3}$\,\cite{Mei2016:impactneutralimpurity}. Neutral impurity scattering does not depend on the temperature: $\mu_{e}^{neu} \propto \frac{m^*_{e}}{\rho_{neu}}$\,\cite{Erginsoy1950:NeutralImpurity}.
    \item \textbf{Lattice dislocations, $dis$:} Imperfections in the crystal lattice can also act as scattering centers. The dislocation concentration is defined as the total length of dislocation lines in a volume and has, hence, the dimension of m$^{-2}$.  In germanium detectors, typical concentrations are of the order of $\rho_{dis} = 10^2 -  10^4$\,cm$^{-2}$\,\cite{Wang2015:Highpuritygermanium}. This is the level required for the detectors to work. The effects of lattice dislocations become relevant at a defect density of $\rho_{dis} > 10^7$\,cm$^{-2}$. Thus, they can be neglected for typical germanium detectors\,\cite{Mei2016:impactneutralimpurity,SeitzTurnbull1958:SSP}.
\end{itemize}
All contributions overlap and are difficult to study separately. The dominant contribution to the temperature dependence of $\mu_{e}$ for germanium detectors at typical operational temperatures, i.e.~70--80\,K, is predicted to be the scattering off acoustic phonons\,\cite{Bardeen1950:DeformationPotentialsMobilities}. The temperature dependence of $\mu_{e}$ is, therefore, expected to be $\mu_{e}(T) \propto T^{-3/2}$. Previous measurements\,\cite{Abt2011:MeasurementTemperatureDependence}, however, indicate that this theoretical prediction is not accurate and that the temperature dependence of the charge carrier drift in germanium should be studied in more detail.
\section{The Detector and Experimental Setup}
\begin{figure}[htb]
    \centering
    \vspace{0.5cm}
  \subfloat[Top view.]{\begin{overpic}[width = 0.48 \textwidth,,tics=10]
        {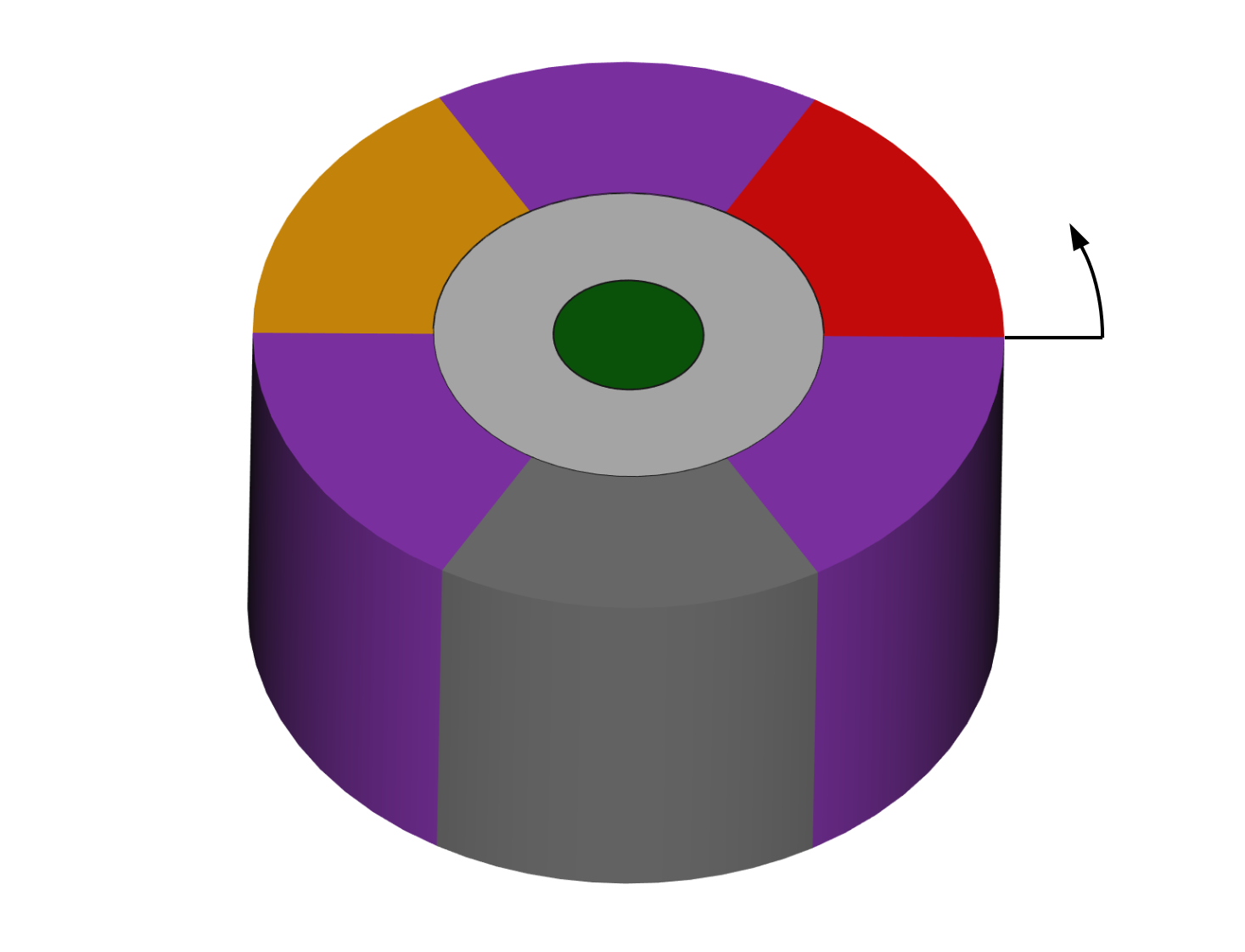}
        \thicklines
        \put(80,63,10) {$\varphi = 0\,^{\circ}$}
        \put(47.5,50){\color{white}\vector(1,0){9}}
        \put(53.5,50){\color{white}\vector(-1,0){9}}
        \put(43,44) {\color{white}15\,mm}

        \put(38.5,40){\color{white}\vector(1,0){26.5}}
        \put(62.0,40){\color{white}\vector(-1,0){26.5}}
        \put(43,35) {\color{white}39\,mm}

        \put(17.0,28){\color{black}\vector(0,1){22.0}}
        \put(17.0,47){\color{black}\vector(0,-1){22.0}}
        \put(11,29) {\rotatebox{90}{40.4\,mm}}

        \put(23.0,3){\color{black}\vector(1,0){57.0}}
        \put(77.0,3){\color{black}\vector(-1,0){57.0}}
        \put(41,-3) {75.05\,mm}

        \put(20.0,68.0){\color{black}\vector(2,-1){30.0}}
        \put(8,68.0){Core}

        \put(84.0,25){\color{black}\vector(0,1){12.0}}
        \put(84.0,25){\color{black}\vector(1,0){12.0}}
        \put(86,37) {$z$}
        \put(96,20) {$r$}

    \end{overpic}
    \vphantom{\includegraphics[width=0.05\textwidth, valign = t]{Figures/Ch_3/segBEGe_top_cor_w_arrow.png}}
    }
    \subfloat[Bottom view.]{
    \begin{overpic}[width = 0.48  \textwidth,,tics=10]
        {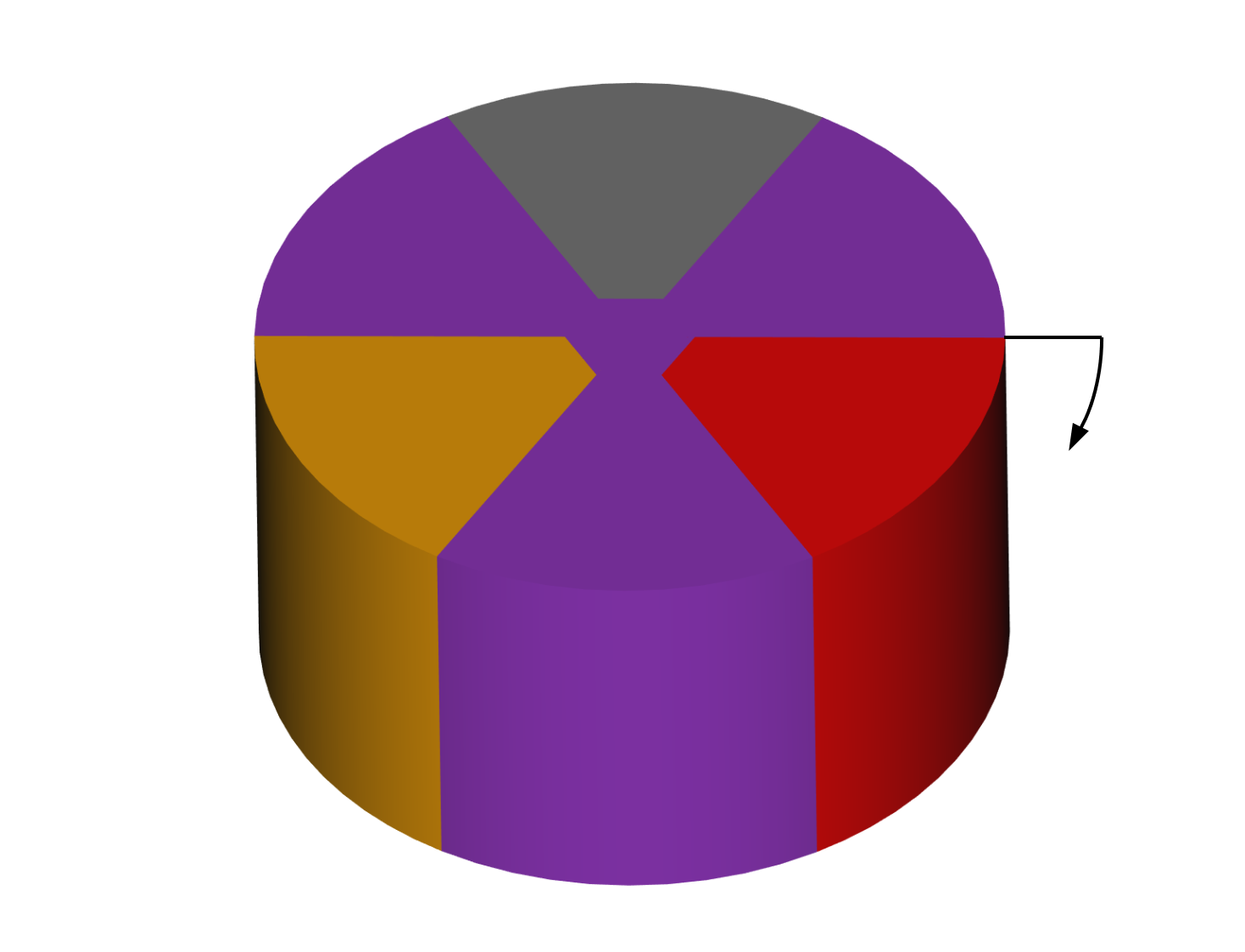}
        \thicklines
        \put(83,53,10) {$\varphi = 0\,^{\circ}$}
        \put(12,66){\color{black}\vector(2,-1){12.0}}
        \put(4,68){Seg.\,4}
        \put(75,75){\color{black}\vector(-2,-1){12.0}}
        \put(71,77){Seg.\,3}
        \put(12,9){\color{black}\vector(2,1){12.0}}
        \put(4,4){Seg.\,2}
        \put(90,9){\color{black}\vector(-2,1){12.0}}
        \put(83,4){Seg.\,1}

        \put(16.0,50){\color{black}\vector(0,-1){15.0}}
        \put(16.0,50){\color{black}\vector(-1,0){15.0}}
        \put(11.0,36) {$z$}
        \put(2,52) {$r$}

        \put(51,74){\color{black}\vector(0,-1){25.0}}
        \put(43,77) {(0,0,0)}
    \end{overpic}
        \vphantom{\includegraphics[width=0.05\textwidth, valign = t]{Figures/Ch_3/segBEGe_top_cor_w_arrow.png}}}
    \caption[Schematic of the segmented BEGe detector.]{Schematic of the segmented BEGe detector. The five read-out channels include the so-called "Core" as well as four segments sharing the mantle surface. Segment 4 covers roughly three times more area than each of the other three equally sized segments. This schematic also defines the cylindrical detector coordinate system used for the remainder of this paper.}
    \label{fig:segBEGe}
    \vspace{0.5cm}
\end{figure}

An experimental setup to create energy deposits close to the surface of a detector was built at the Max Planck Institute for Physics in Munich to specifically study charge drift\,\cite{PhD:Schuster2021}. 
The detector under study was an n-type segmented BEGe\footnote{BEGe stands for "Broad Energy Range Germanium" and is the name for a product series of point-contact detectors of the manufacturer MIRION France. The detector presented here is a customized prototype.} detector\,(segBEGe).
The geometry and the layout of the contacts of the segBEGe are shown in Fig.\,\ref{fig:segBEGe}.\\
Around the point contact, which is commonly referred to as the "Core", there is a passivated ring which protects the non-contacted surface. The rest of the detector surface is divided into four segments, three of which are equally sized and reach over 60$^{\circ}$ across the top, side and bottom of the detector. They are located every 120$^{\circ}$ and are referred to as Segments 1, 2 and 3. The remaining surface in-between is covered by Segment 4, which is closed at the bottom end-plate. Figure~\ref{fig:segBEGe} also introduces the cylindrical coordinate system which will be the reference frame for all analyses. The origin is located at the bottom center of the detector, with the $z$-axis pointing upwards towards the Core and the azimuthal axis starting with $\varphi = 0^{\circ}$ at the left border of Segment 1, when looking from the side. This configuration provides a total of five read-out channels each of which provides an individual signal for a given charge drift.\\

The segBEGe is operated with a $U_{RB}$ of $+4500$\,V applied to the Core. The electrons are collected at the Core while the holes are collected at the segment electrodes. Table~\ref{tab:segBEGe} provides an overview on the relevant specifications of the crystal as provided by the manufacturer.\\
\begin{table}[htb]
    \centering
\caption{Specifications of the segmented BEGe detector as provided by the manufacturer.}\label{tab:segBEGe}
    \begin{tabular}{l|rr}
        \tbf{Height}  & 40.4 & mm\\
        \tbf{Crystal diameter}  & 75.05 & mm\\
        \tbf{Core contact diameter} & 15.0 & mm \\
        \tbf{Passivation ring  diameter} & 39.0 & mm \\
        \tbf{Impurity concentration (top)} & 1.30 & $\cdot$\,10$^{10}$\,cm$^{-3}$\\
        \tbf{Impurity concentration (bottom)} & 0.95 & $\cdot$\,10$^{10}$\,cm$^{-3}$\\
        \tbf{Recommended} $\bm{U_{RB}}$ & +\,4500 &  V 
    \end{tabular}
    \vspace{0.4cm}
\end{table}%
\FloatBarrier
The segBEGe was housed in the K2 Cryostat, a multi-purpose electrically cooled cryostat from the manufacturer MIRION France\,\cite{PhD:Schuster2021}. An adjustable and stable temperature of the segBEGe was essential for the study presented here. K2 allowed stable running at temperatures down to $\approx 73$\,K. The temperature was read out at different points in the cryostat via a total of four PT-100 temperature sensors. Since there was no PT-100 sensor directly on the crystal, the average of the two temperatures measured closest to the crystal, i.e.~onto the stage on which the crystal was mounted, is taken as the crystal temperature, $T_{det}$. The actual crystal temperature could be up to 2\,K higher. However, this shift is constant to $\pm 0.1$\,K for a given setup\,\cite{Canberra:PrivateCommunication2}. A designated aluminum ring around the center of the cryostat houses the preamplifier boards for all read-out channels.\\
A set of components was installed around the K2 cryostat, which allowed to move a collimated source around the detector and scan it in a precise and reproducible way. The setup consisted of a frame of MayTec aluminum profiles mounted on a rotation stage controlled by a STANDA stepper motor
with a precision of 0.15$^{\circ}$, which corresponds to 0.1\,mm at a radius of 37.5\,mm. A linear motor stage
which was precise to 0.005\,mm was mounted on a vertical aluminum profile, holding a collimated $^{133}$Ba source. The two motors were controlled remotely. Software was developed to scan the side of the segBEGe repeatedly in a completely automated way.\\
The source isotope $^{133}$Ba was chosen as it features a characteristic low energy gamma line of 81\,keV\cite{def:Firestone1999}. 
The mean free path of 81\,keV gammas in germanium is 2.68\,mm\,\cite{MEAC}. Monte-Carlo simulations of the mean penetration depth yielded a compatible value of 2.50\,mm\,\cite{PhD:Schuster2021}.\footnote{This value is used from here on.} The size of the beam spot on the detector surface of 2.3\,mm was also determined from these simulations.\\  

\begin{figure}[htb]
\centering
\includegraphics[width = \textwidth]{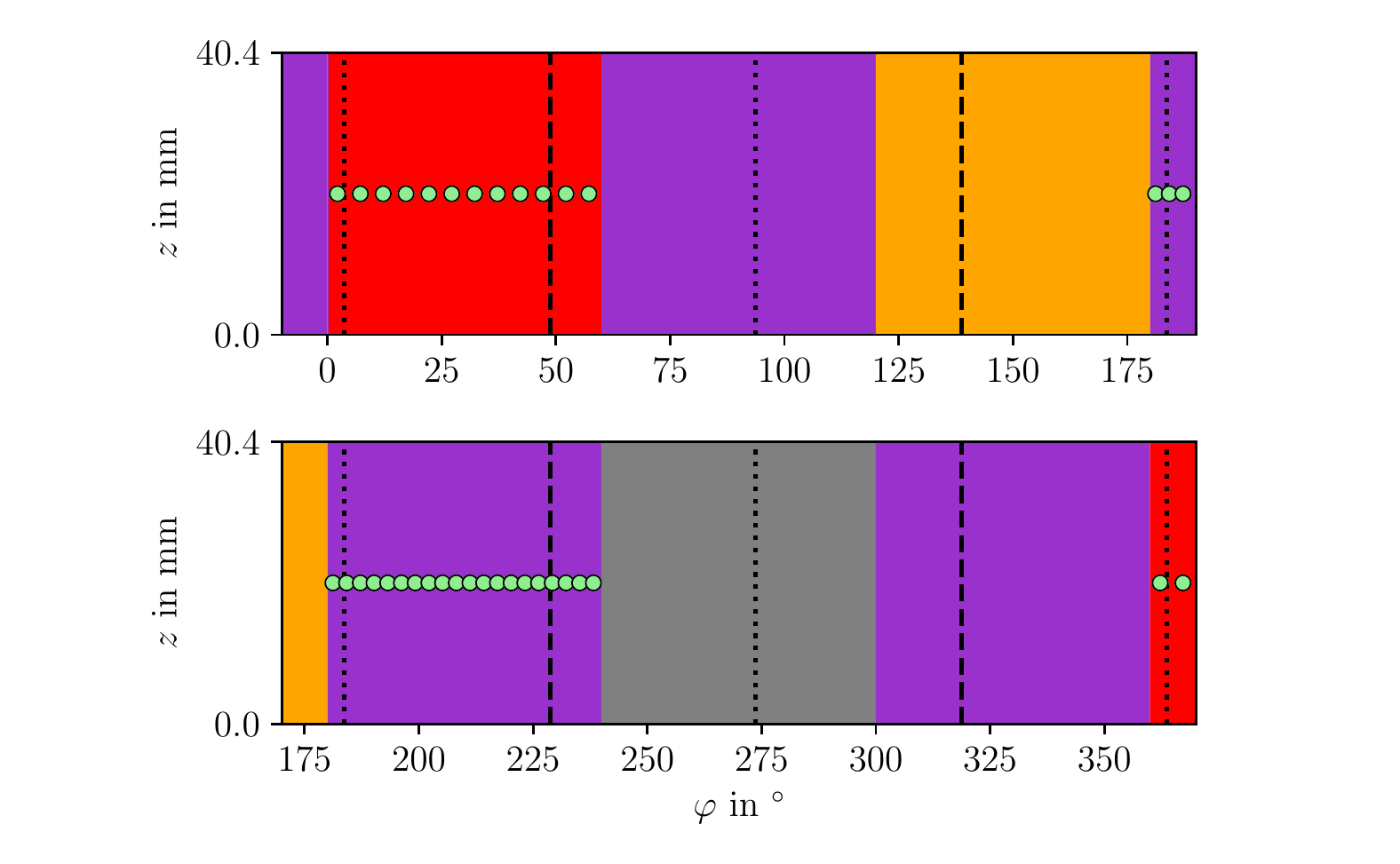}
\caption{Schematic of the side of the segBEGe using the color scheme introduced in Fig.\,\ref{fig:segBEGe}. The dots indicate the positions at which the detector was irradiated. The size of the dots reflects the size of the beam spot on the detector surface, i.e.~a 2.3\,mm diameter. The dotted\,(dashed) vertical lines indicate the positions of the $\langle110\rangle$\,($\langle100\rangle$) axes.}\label{fig:data_taking}
\end{figure}

The data were taken using a 14-bit analog-to-digital converter\,(STRUCK SIS3316-250-14) with a sampling frequency of 250\,MHz, i.e.~a sampling time of 4\,ns. The $^{133}$Ba source had an activity of $\approx 1$\,MBq and produced a large number of near-surface events with a signal-to-background ratio $\geq$\,10 for the 81\,keV line. For 15 temperatures across the range from $\approx 73-118$\,K, two azimuthal scans along the side of the detector at $z = 20.2$\,mm were taken:~one across Segment~1, called AS-1, and the other across the part of Segment 4 which is located between Segments 2 and 3, called AS-4. For AS-1\,(AS-4) a step size of 5\,$^{\circ}$\,(3\,$^{\circ}$) was chosen. Each point was irradiated for 10 minutes. Figure \ref{fig:data_taking} shows a schematic of the side of segBEGe together with the positions of the beam spot during both scans.\\

\begin{figure}[htb]
    \centering
    \begin{overpic}[width=\textwidth]
    {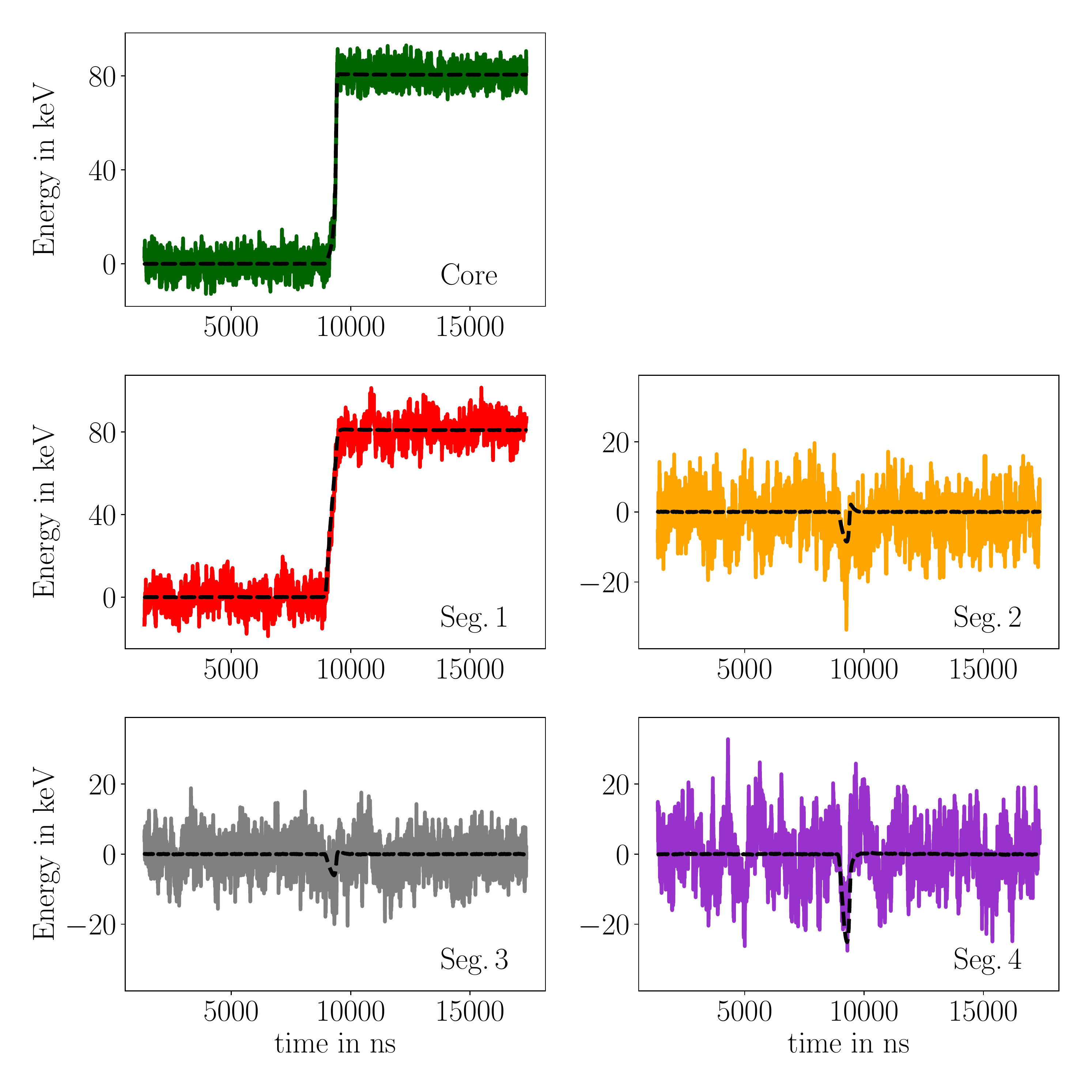}
        \put(60,94){\textbf{DATA}}
        \put(60,89){segBEGe}
        \put(60,84){$^{133}$Ba}
        \put(60,79){$T_{det}$\,=\,77.0\,K}
        \put(60,74){$\varphi$\,=\,47.2$\,^{\circ}$, $z$\,=\,20.2\,mm }
    \end{overpic}
    \caption{Pulses for all read-out channels for a typical 81\,keV event after energy calibration and cross-talk correction. Segment 1 is the collecting segment. The dashed black lines are the 81\,keV superpulses for each channel, obtained by averaging pulses from 16720 selected events.}
    \label{fig:example_pulse}
\end{figure}
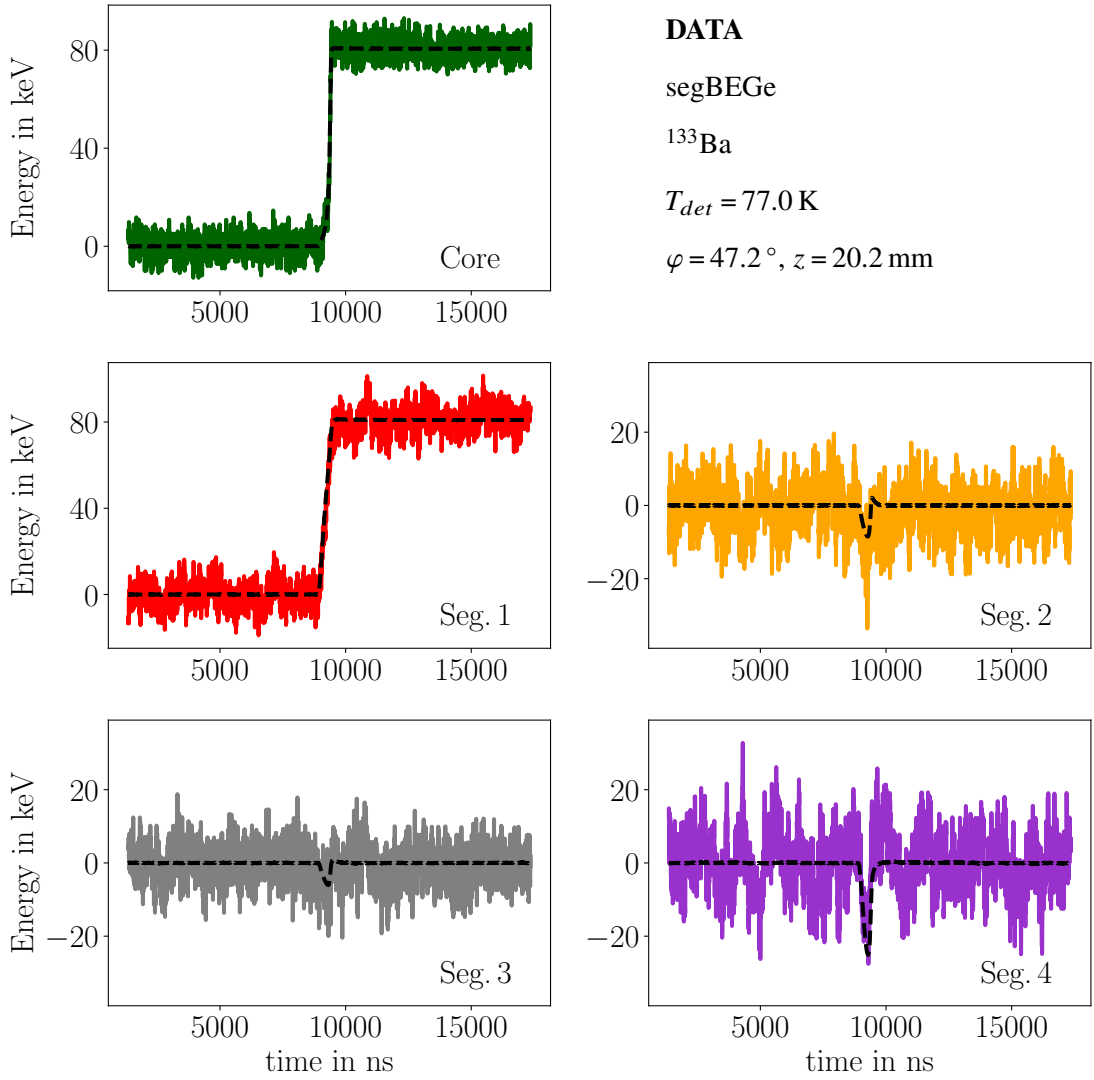

The resulting pulses were calibrated and corrected for linear cross-talk offline\,\cite{PhD:Schuster2021}. Pulses corresponding to the same scan location were averaged to form so-called superpulses to mitigate the effects of electronic noise. Only pulses from events passing a set of selection criteria contributed to the superpulses: The energy registered in the Core had to be within 81\,$\pm\,2$\,keV, the energy registered in the collecting segment had to be within 2\,keV of the Core energy, and the event had to pass pile-up and data quality cuts\,\cite{PhD:Schuster2021}.\\
A typical 81\,keV event from a $^{133}$Ba measurement is shown in Fig.\,\ref{fig:example_pulse}. Also shown are the corresponding superpulses which reveal pulse shape features otherwise obstructed by the electronic noise, especially for the non-collecting segments. These superpulses are the basis of the analyses presented here. The positions of the crystal axes relative to the segment boundaries were determined previously\,\cite{PhD:Schuster2021,Abt2019:Characterization}.
\FloatBarrier
\section{Observed Temperature Dependence of Rise Times}\label{sec:Tdep-rt5-95}
The time it takes a pulse to rise to its full amplitude is called the rise time, $t_{rt}$. It is, to first order, inversely proportional to the drift velocity, which itself is different along different crystallographic axes, see Sec.\,\ref{sec:ccdrift_ge}. Thus, the rise time depends on the course of drift trajectory with respect to the crystal axes and on the total length of the drift path, and is a quantity which e.g.~can be used to find the crystallographic axes in a detector.\\
The rise time is typically defined for a start and end percentage with respect to the full pulse amplitude, e.g.~5--95\%, denoted as $t_{rt}^{5-95}$. For the AS-1 and AS-4 data, the electrons drift inwards horizontally for some distance, before the drift paths bend upwards towards the Core. The holes drift a small distance towards the mantle. Therefore, $t^{5-95}_{rt}$ represents a superposition of the electron drift along different axes and a small contribution from the hole drift. The uncertainties on $t_{rt}^{5-95}$, $\sigma_{rt}$, were estimated by forming auxiliary superpulses from subsets of the selected pulses used to form the respective superpulses\,\cite{PhD:Schuster2021}. A typical uncertainty of ${\sigma}_{rt} = 3$\,ns on $t_{rt}^{5-95}$ was found and assigned to all rise times.\\

\begin{figure}[htb]
    \centering
    \includegraphics[width = .9\textwidth]{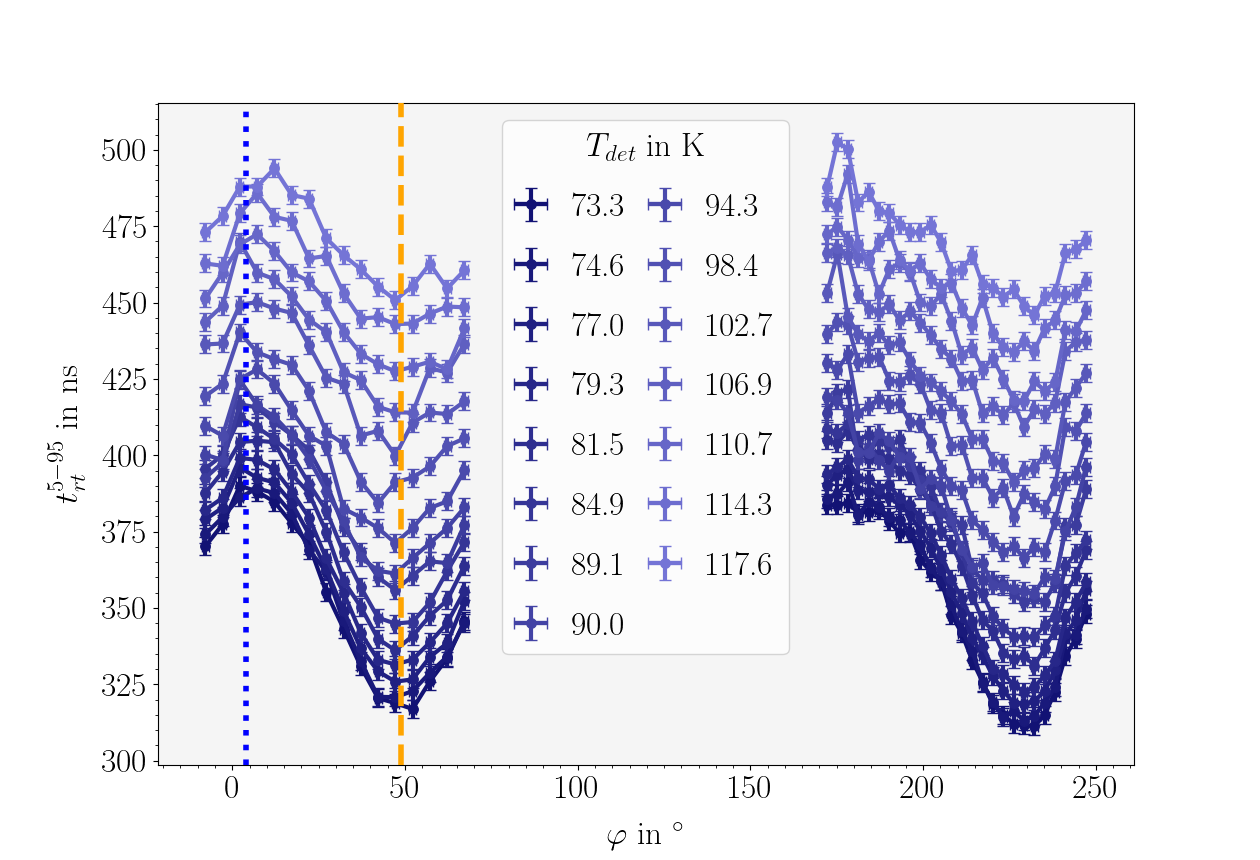}
    \caption{Rise times $t_{rt}^{5-95}$ determined from 81 keV Core superpulses versus $\varphi$. The lines between the points were added to guide the eye. The dotted\,(dashed) line indicates the location of the \br{110}\,(\br{100}) axis as previously determined\,\cite{PhD:Schuster2021}. The vertical error bars represent a 3\,ns uncertainty on $t_{rt}^{5-95}$.\label{fig:risetimes_over_phi_5-95}}
\end{figure}

\begin{figure}[htb]
    \centering
    \includegraphics[width = .9\textwidth]{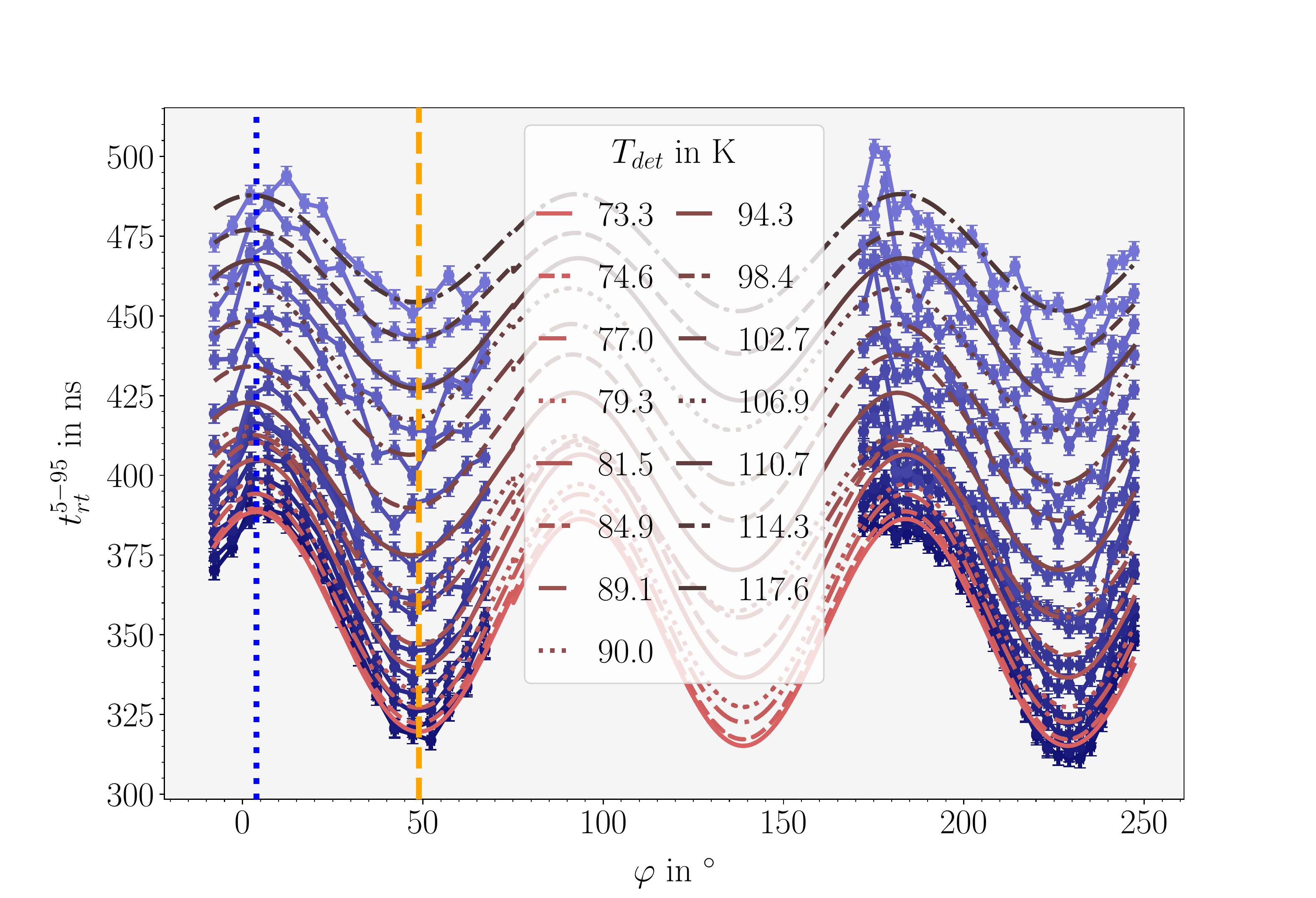}
    \caption{Rise times $t_{rt}^{5-95}$ determined from 81\,keV Core superpulses versus $\varphi$ as in Fig.\,\ref{fig:risetimes_over_phi_5-95} together with fits according to Eq.\,\eqref{eq:modified_sine}. Other details as in Fig.\,\ref{fig:risetimes_over_phi_5-95} .\label{fig:risetimes_over_phi_5-95_w_fits}}
\end{figure}

The $t_{rt}^{5-95}$ for the Core for all data points of AS-1 and AS-4 and for all temperatures are shown  versus $\varphi$ in Fig.\,\ref{fig:risetimes_over_phi_5-95}. The 90$^{\circ}$ oscillation pattern reflecting the location of the crystallographic axes is present for all temperatures. Two major observations are that the $t_{rt}^{5-95}$ are significantly longer for higher $T_{det}$ and that the difference in $t_{rt}^{5-95}$ for the different crystal axes, i.e.~the amplitude of the cosine-like pattern, decreases for higher $T_{det}$.\\
The individual points are subject to fluctuations which are larger for higher temperatures due to increased noise. In order to extract consistent temperature dependences for individual $\varphi$ values, a cosine function, allowing for shifts in the offset and amplitude between the AS-1 and AS-4 data blocks, was fitted to the data for each $T_{det}$ individually, i.e.
\begin{linenomath}
\begin{equation}\label{eq:modified_sine}
    f(\varphi) = p^{i}_{1} +  p^{i}_{2} \cdot \cos\left(\dfrac{2\pi}{90^{\circ}} \cdot (\varphi - p_{3}) \right)~~~~~,
\end{equation}
\end{linenomath}
where $p^{i}_{1}$ denote the constant offsets and $p^{i}_{2}$ are the amplitudes for the two blocks with $i \in  \{$AS-1,\,AS-4$\}$. The offset in $\varphi$, $p_{3}$, is a common parameter for AS-1 and AS-4. Figure~\ref{fig:risetimes_over_phi_5-95_w_fits} shows the resulting fit functions, $f_{t_{rt}^{5-95}}$, together with the data.\\

\begin{figure}[htb]
    \centering
    \begin{overpic}[width = .9\textwidth]{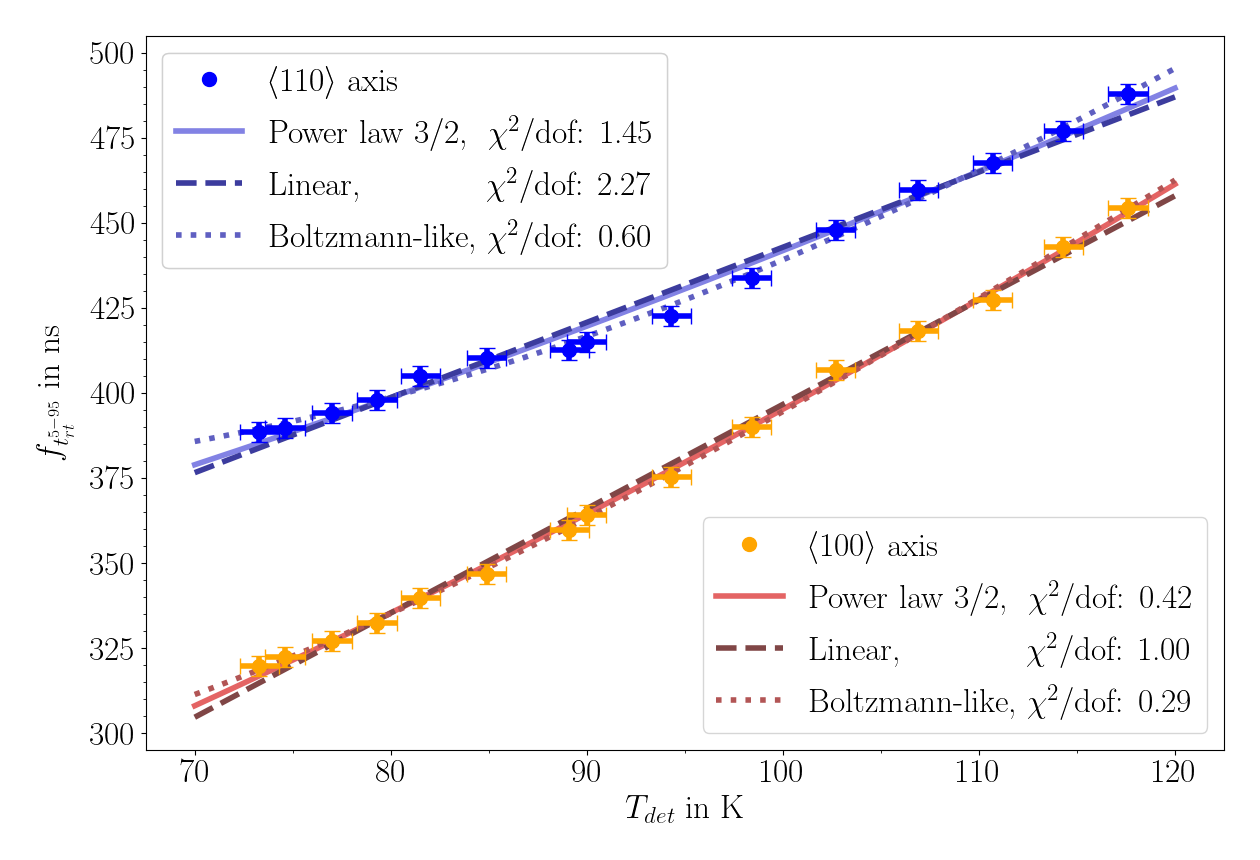}
    \put(19,17) {\vector(0,1){13}}
    \put(19,30) {\vector(0,-1){13}}
    \put(21,23.5) {\textbf{21.6\,\%}}
    
    \put(91,53.5) {\vector(0,1){5}}
    \put(91,58.5) {\vector(0,-1){5}}
    \put(88,45) {\textbf{7.3\,\%}}
    \end{overpic}
    \caption{Rise times $f_{t_{rt}^{5-95}}$ as determined from the fits, see Eq.\,\eqref{eq:modified_sine} and Fig.\,\ref{fig:risetimes_over_phi_5-95_w_fits}, for both major axes in Seg.\,1, i.e.~$\varphi\,=\,48.87\,^{\circ}$ for $\langle100\rangle$ and $\varphi\,=\,3.87\,^{\circ}$ for $\langle110\rangle$, versus $T_{det}$. Also shown are fits according to Eqs.\,\eqref{eq:powerlaw32}--\eqref{eq:boltzmannlike}.}\label{fig:risetimes_over_T_5-95}
\end{figure}
The values for $f_{t_{rt}^{5-95}}(\varphi)$ versus $T_{det}$ are shown in Fig.\,\ref{fig:risetimes_over_T_5-95} for the $\varphi$ corresponding to the $\langle110\rangle$ and $\langle100\rangle$ axes positions in Seg.\,1. Over the observed temperature range of $\approx 45$\,K, $f_{t_{rt}^{5-95}}$ increases significantly for both axes. For electrons starting their drift along the \br{100} axis, $f_{t_{rt}^{5-95}}$ increases faster with increasing $T_{det}$ than for electrons starting their drift on the \br{110} axis. Thus, the difference in $f_{t_{rt}^{5-95}}$ decreases with rising $T_{det}$. Three model functions were fit to $f_{t_{rt}^{5-95}}$ for both axes using \lsq{}\,\cite{web:LsqFit.jl}:
\begin{itemize}
  \item
  \textbf{Power law dependence:} 
  As pointed out in Sec.\,\ref{sec:tdep}, the predominant contribution $\mu_{e}$ in the relevant temperature range is expected to be $\mu_{e}^{aco}$, i.e.~the scattering off acoustic phonons. Hence, $\mu_{e}$ is expected to scale with $T^{-3/2}$. As $t_{rt} \propto \frac{1}{\mu_{e}}$, a power-law function with a power of $3/2$ is a natural choice as a model function. Earlier results\,\cite{Abt2011:MeasurementTemperatureDependence} indicate, however, that this $ansatz$ does not describe the temperature dependence correctly. Therefore, an additional constant offset, which could arise due to the presence of neutral impurities, $\frac{1}{\mu_{e}(T)} \approx \frac{1}{\mu_{e}^{aco}(T)} + \frac{1}{\mu_{e}^{neu}}$ was introduced as a free parameter in an extension of the model:
  \begin{linenomath}
  \begin{equation}\label{eq:powerlaw32}
    F^{pl}(T) = p_0 + p_1 \cdot  T^{3/2}~~~~~.
  \end{equation}
  \end{linenomath}
  \item
  \textbf{Linear dependence:}
  As a simple reference model, a linear dependence on $T$ was also considered:
      \begin{equation}\label{eq:linear}
      F^{lin}(T) = p_0 + p_1 \cdot T~~~~~.
    \end{equation}
  \item
  \textbf{Boltzmann-like dependence:}
In a previous publication\,\cite{Abt2011:MeasurementTemperatureDependence}, a Boltzmann-like function was found to describe the data of an n-type true-coaxial germanium detector best:
  \begin{linenomath}
  \begin{equation}\label{eq:boltzmannlike}
      F^{bm}(T) = p_0 + p_1 \cdot \exp(-p_2 / T)~~~~~.
  \end{equation}
  \end{linenomath}
\end{itemize}
The $\chi^2/$degrees of freedom\,(dof) values for a given $\varphi$ were determined as
\begin{linenomath}
\begin{equation}
\chi^2/\text{dof} =\left( \sum\limits_i^{N_T = 15} \dfrac{(~F^{*}(T_{det,i}) - f(T_{det,i}|\varphi)~)^2}{\sigma_{rt}^2}\right)~/~\text{dof}~~~~~,
\end{equation}
\end{linenomath}
where $F^*$ is the respective model function,  $\sigma_{rt}\,=\,3$\,ns~ and $N_T$ is the total number of the different temperatures, $T_{det,i}$. The number of dof is $N_T$ minus the number of parameters of the respective fit function $F^*$. The uncertainty on $T_{det,i}$ is systematic and was not taken into account in the fits. To investigate the influence of this uncertainty, the fits were also performed for the data shifted by $\pm\,1$\,K. The results of the fits of Eqs.\,\eqref{eq:powerlaw32}--\eqref{eq:boltzmannlike} to the data are shown in Fig.\,\ref{fig:risetimes_over_T_5-95} and summarized in Tab.\,\ref{tab:n-type_rt_vs_T_results}.\\

\begin{table}[htb]
\renewcommand*{\arraystretch}{1.2} 
  \centering
  \caption[Results of the fits of three model functions to rise times of side scan superpulses against temperature for the n-type segmented BEGe.]{Results of the fits according to Eqs.\,\eqref{eq:powerlaw32}--\eqref{eq:boltzmannlike} as shown in Fig.\,\ref{fig:risetimes_over_T_5-95}. The experimental uncertainty refers to the 90\% confidence interval of the respective fit parameter as determined with \lsq{}.
The systematic uncertainties result from fits where $T_{det}$ was shifted by $\pm$\,1\,K. Where no systematic uncertainty is given, it was $<0.01$.}\label{tab:n-type_rt_vs_T_results}
   \vspace{0.2cm}
   \resizebox{\textwidth}{!}{%
  \begin{tabular}{l|c|ccc|r}
      & \textbf{Axis} & $\bm{p_0}$\,$\bm{\pm}$\textbf{(exp)}\,$\bm{^{+}_{-}}$\textbf{(syst)} & $\bm{p_1}$\,$\bm{\pm}$\textbf{(exp)}\,$\bm{^{+}_{-}}$\textbf{(syst)} & $\bm{p_2}$\,$\bm{\pm}$\textbf{(exp)}\,$\bm{^{+}_{-}}$\textbf{(syst)} & $\bm{\chi^2}$\textbf{/dof} \\
      \hline
      \multirow{2}{*}{\textbf{Eq.}\,\eqref{eq:powerlaw32}} & $\langle100\rangle$ & 184.8\,$\pm6.0$\,$^{+2.0}_{-2.0}$ & 0.21\,$\pm0.01$ & / & 0.42\\
      & $\langle110\rangle$ & 289.9\,$\pm6.0$\,$^{+1.5}_{-1.5}$& 0.15\,$\pm0.01$ & /& 1.45\\
      \hline
    \multirow{2}{*}{\textbf{Eq.}\,\eqref{eq:linear}} & $\langle100\rangle$ & 90.0\,$\pm8.8$\,$^{+3.1}_{-3.1}$ & 3.07\,$\pm0.09$ & / & 1.00\\
      & $\langle110\rangle$ & 221.9\,$\pm8.8$\,$^{+2.2}_{-2.2}$ & 2.21\,$\pm0.09$ & / & 2.27 \\
      \hline
      \multirow{2}{*}{\textbf{Eq.}\,\eqref{eq:boltzmannlike}} & $\langle100\rangle$ & 272.0\,$\pm17.8$\,$^{+0.7}_{-0.7}$ & 1734\,$\pm560$\,$^{+28.}_{-28}$ & 264.9\,$\pm47.6$\,$^{+3.7}_{-3.7}$ & 0.29\\
        & $\langle110\rangle$ & 371.5\,$\pm9.9$\,$^{+0.3}_{-0.3}$ & 2556\,$\pm1347$\,$^{+64}_{-62}$ & 363.2\,$\pm69.0$\,$^{+5.7}_{-5.7}$ & 0.60  \\
        \hline
        \hline
    \end{tabular}
    }
     \vspace{0.5cm}
\end{table}%
In general the Boltzmann-like function gives the best fits. However, all three models describe the data reasonably well. The $t_{rt}^{5-95}$ as discussed in the section can only be used for very general statements, because the electrons drift along and between different axes, before reaching the Core. In order to extract information on individual axes, help from simulation is needed.
\FloatBarrier
\clearpage
\section{Simulation and Custom Rise-Time Windows}
\begin{wrapfigure}{r}{0.40\textwidth}
    \centering
            \begin{overpic}[width = .40\textwidth,,tics = 10]
            {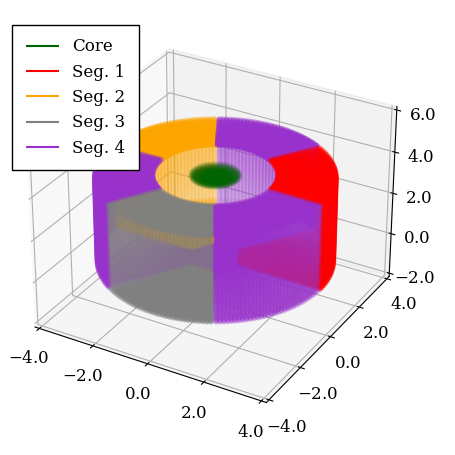}
            \put (80,10) {\small\rotatebox{47}{$y$\,in cm}}
            \put (14,9) {\small\rotatebox{-22}{$x$\,in cm}}
            \put (98,50) {\small\rotatebox{90}{$z$\,in cm}}
        \end{overpic}
    \caption{Visualization of the implementation of the segBEGe in \SSD{}.}
    \label{fig:segBEGe_SSD}
\end{wrapfigure}
\begin{figure}[tb]
    \centering
    \begin{overpic}[width = \textwidth, , tics = 10]{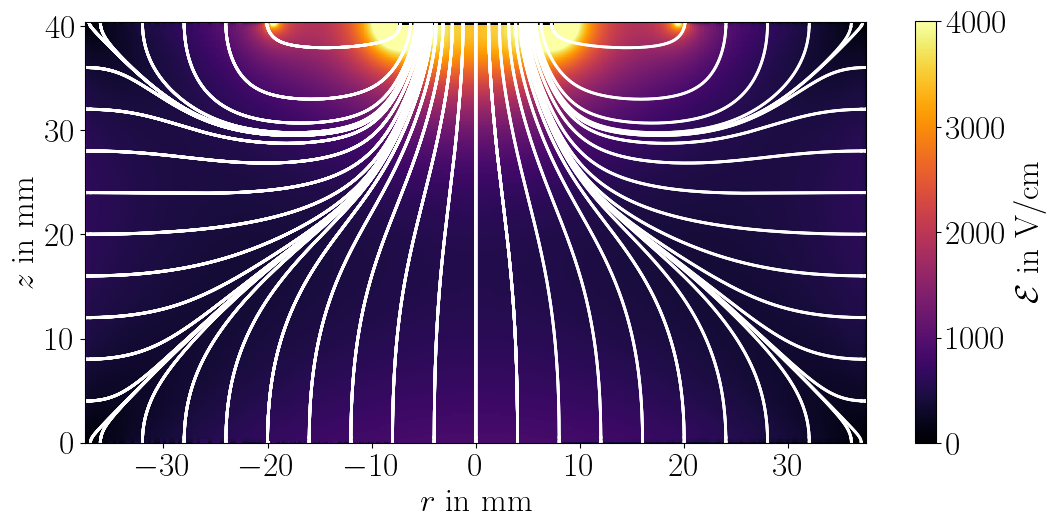}
    \put (5,1) { $\bm{\varphi = 30^{\circ}\,(210^{\circ})}$} 
    \end{overpic}
    \caption{Electric-field strength along $r$ and $z$, as calculated with \SSD{}. The electric-field lines are indicated in white. The color bar is artificially capped at 4000\,V/cm to diminish grid-related artifacts and reveal the field gradients in the bulk. Negative radii correspond to $\varphi = 210^{\circ}$.}
    \label{fig:electric_field}
\end{figure}

%
For all simulations of the charge carrier drift and the resulting pulses, the open-source software package \SSD{}\,\cite{SSD2021:SSDPaper}, $SSD$, was used at version v0.7.3. Figure~\ref{fig:segBEGe_SSD} shows the implementation of the segBEGe. The electric potential is calculated on an adaptive grid, which is refined in high-gradient regions.\\
The impurity density profile of the crystal is an input to the simulation. It was also used to calculate the detector capacitance vs.~$U_{RB}$. The manufacturer provided measurements of the capacitance of the detector vs.~$U_{RB}$ and the impurities. However, the impurities are less well determined and according to the calculation they did not correspond well to the measured capacitances. Comparisons between simulated and provided capacitance values lead to a down-scaling of the impurities to 35\,\% of the values given in Tab.\,\ref{tab:segBEGe}.\\
Figure\,\ref{fig:electric_field} shows $\mathcal{E}$ in the $r$-$z$-plane at $\varphi = 30^{\circ}$ as calculated with $SSD$. For $z \approx 20$\,mm, the $\mathcal{E}$ values decline with smaller radii, from $\lessapprox 1000$\,V/cm at the surface to a few hundred towards the center of the detector. Strong fields, $\mathcal{E} > 3000$\,V/cm are only observed in the semi-ellipsoidal volume around the Core contact. The charges drift, to first order, along the electric-field lines, indicated in white.\\

The drift-velocity vectors for electrons and holes are calculated based on the electric-field vectors, which are determined at every grid point. Both, for electrons and holes, the longitudinal drift velocities along the $\langle100\rangle$ and $\langle111\rangle$ axes are calculated according to Eq.\,\eqref{eq:drift_velocity} with the parameters determined from measurements at $T_{det}\,=\,78$\,K\,\cite{Bruyneel2006:CharacterizationII}, see Tab.\,\ref{tab:mobilities}. From the velocities along \br{100} and \br{111}, the velocities in all directions are calculated using the respective charge drift model for electrons and holes. Both charge drift models are implemented as described in Ref.\,\cite{SSD2021:SSDPaper}. The electron model is summarized in App.\,\ref{app:e_drift_model}.\\
The relation between the longitudinal velocity along \br{110} and the longitudinal velocities along \br{100} and \br{111} is linear \,\cite{Mihailescu:2000jg, Hagemann19:MT}.
For electrons,
\begin{linenomath}
\begin{equation}\label{eq:ve110}
    v_{e}^{110} = \beta_{100} \cdot v_{e}^{100} + \beta_{111} \cdot v_{e}^{111}~~~~~,
\end{equation}
\end{linenomath}
where $\beta_{100}\,=\,0.786$ and $\beta_{111}\,=\,0.200$. A linear relation is also found for holes:
\begin{linenomath}
\begin{equation}\label{eq:vh110}
        v_{h}^{110} = \frac{1}{4} \cdot v_{h}^{100} + \frac{3}{4} \cdot v^{111}_{h}~~~~~.
\end{equation}
\end{linenomath}

The temperature dependence is implemented in $SSD$ by scaling $v_{e/h}^{\langle100\rangle}$ and $v_{e/h}^{\langle111\rangle}$ by factors calculated according to the temperature-dependence model chosen by the user, i.e.~the user has to provide a model function and its parameters.\footnote{Usually, there are two different sets of parameters for the two different axes.} This approach corresponds to an effective scaling of $\mu_{e/h}$\,(and $\mu_{n}$), see Eq.\,\eqref{eq:drift_velocity}.

\begin{figure}[htb]
    \centering
    \begin{overpic}
    [width = .9\textwidth, , tics = 10]{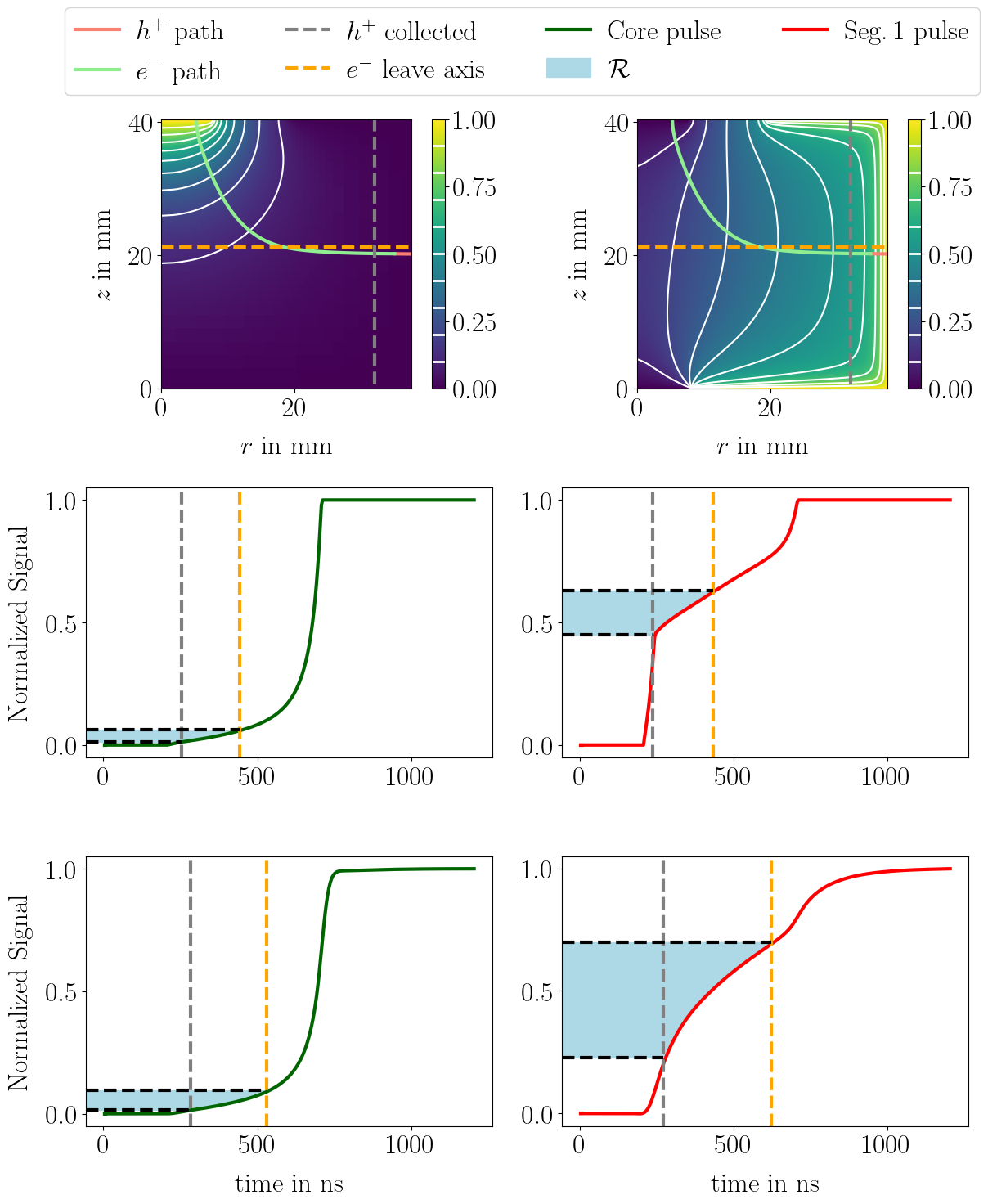}
        \put (0,89) {\textbf{a)}}
        \put(16.2, 68.5) {\small \color{white}$\varphi = 2.2\,^{\circ}$}
        \put(43, 76) {\rotatebox{90}{$\mathcal{W}^{Core}$}}
        \put(82, 76) {\rotatebox{90}{$\mathcal{W}^{Seg.\,1}$}}
        
        \put (0,58) {\textbf{b)}}
        \put (26.5,42) {\textbf{Simulation}}
        \put (8.5, 45) {\textbf{1.1\,-}}
        \put (8.5, 42) {\textbf{5.9\,\%}}
        \put (66, 50) {\textbf{45.0\,-}}
        \put (66, 47) {\textbf{63.1\,\%}}
        
        \put (0,27) {\textbf{c)}}
        \put (26.5,11) {\textbf{Simulation}}
        \put (26.5,8) {\textbf{+ Resp. F.}}
        \put (8.5, 15) {\textbf{1.4\,-}}
        \put (8.5, 12) {\textbf{9.5\,\%}}
        \put (66, 18) {\textbf{22.8\,-}}
        \put (66, 15) {\textbf{69.9\,\%}}
    \end{overpic}
    \caption{a): Cross section of the Core\,(Seg.\,1) weighting potential of the segBEGe on the left\,(right), together with electron and hole, $e^{-}$ and $h^{+}$, trajectories of an 81\,keV event starting in the $r$-$z$-plane at $\varphi=2.2^{\circ}$, close to the $\langle110\rangle$ axis. The white lines are the equipotential lines of $\mathcal{W}^{Core}$\,($\mathcal{W}^{Seg.\,1}$). b): Corresponding simulated pulse for the Core\,(Seg.\,1) on the left\,(right). The dashed lines mark $t_{hc}$ and $t_{et}$. c): Simulated pulse for the Core\,(Seg.\,1) on the left\,(right) after the application of the respective response function. The percentages denote the rise-time windows.}
    \label{fig:crtw_determination}
\end{figure}
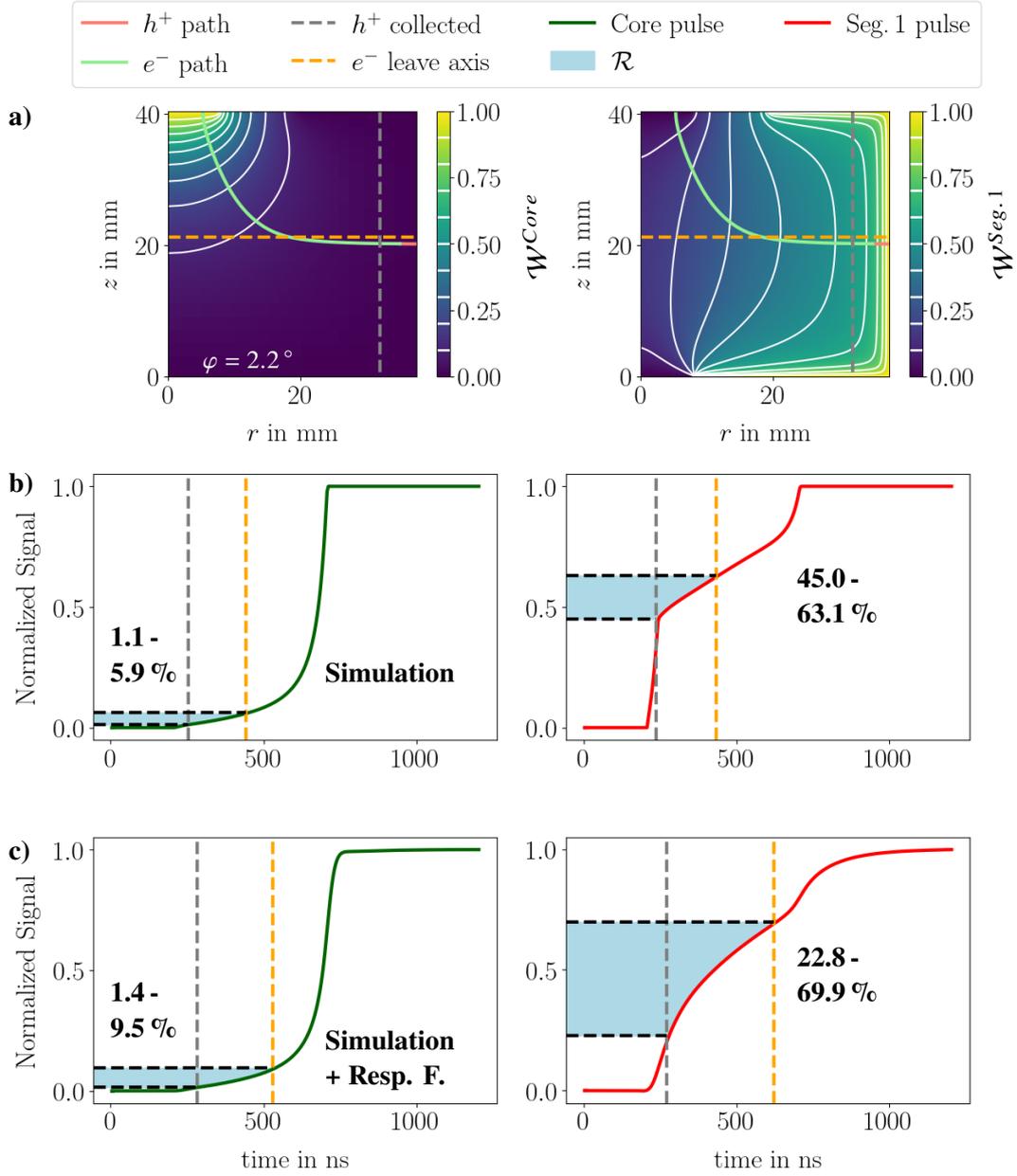

The pulses provided by $SSD$ represent the ideal detector response. Before comparisons to superpulses from data, they were convolved with the corresponding response functions measured for each individual read-out channel\,\cite{PhD:Schuster2021}, see App.\,\ref{sec:response_functions}. \\
The drift of point charges\footnote{It has been investigated that, in this energy range, charge cloud effects such as diffusion and self-repulsion have negligible effects on the pulse shape. Therefore, point charges are a reasonable simplification\,\cite{PhD:Schuster2021}.} from the positions of the AS-1 and AS-4 scans and $r = 35.0$\,mm, corresponding to the mean penetration depth of 81\,keV events, was simulated with $SSD$. The default charge-drift parameters which correspond to a temperature of 78\,K, see Tab.\,\ref{tab:mobilities}, were used. Based on the simulated charge-drift paths, two points of interest were defined and the corresponding timestamps were extracted:
\begin{itemize}
    \item When the holes are collected, $t_{hc}$, and
    \item when the electrons start turning towards the point contact after the initial inwards drift, $t_{et}$.
\end{itemize}
The latter is reached as soon as the current $z$-position of the drifting electrons deviates by 1\,mm from the original $z$-position.\\
The two timestamps define a time window, $t_{hc}<t<t_{et}$, during which only the inwards drift of the electrons contributes to the signal formation. This time window can be translated into a rise-time window, $\mathcal{R}$.
This is illustrated in Fig.\,\ref{fig:crtw_determination} for an AS-1 irradiation point close to the $\langle110\rangle$ axis at $\varphi = 2.2^{\circ}$, at $r = 35.0$\,mm and $z = 20.2$\,mm for the Core and the collecting Seg.\,1. The simulated drift paths of the electrons and holes are shown as a projection to the $r$-$z$-plane at $\varphi = 2.2^{\circ}$, on top of the respective weighting potential, $\mathcal{W}$, which relates the position of the charges to the strength of the induced signal\,\cite{He2001:SRT}. The weighting potential along the electron path between $t_{hc}$ and $t_{et}$ is rather small for the Core channel and significantly larger for Seg.\,1.\\
To use the simulation-based $\mathcal{R}$ for the interpretation of data, the response functions have to be taken into account. The convolution with the respective response function adds the effects of the read-out electronics to the simulated pulses and influences their shape and the total pulse length. Accordingly, $t_{hc}$ and $t_{et}$ were scaled with the increased total pulse length, $t_{rt}^{0.5-99.5}$, while their relative positions with respect to the start and end of the convolved pulse were kept the same.
For both, the Core and Seg.\,1, the corresponding $\mathcal{R}$ are larger after the response functions are taken into account.
\FloatBarrier
\section{Interpretation of Data for Isolated Axes}\label{sec:CRTW}
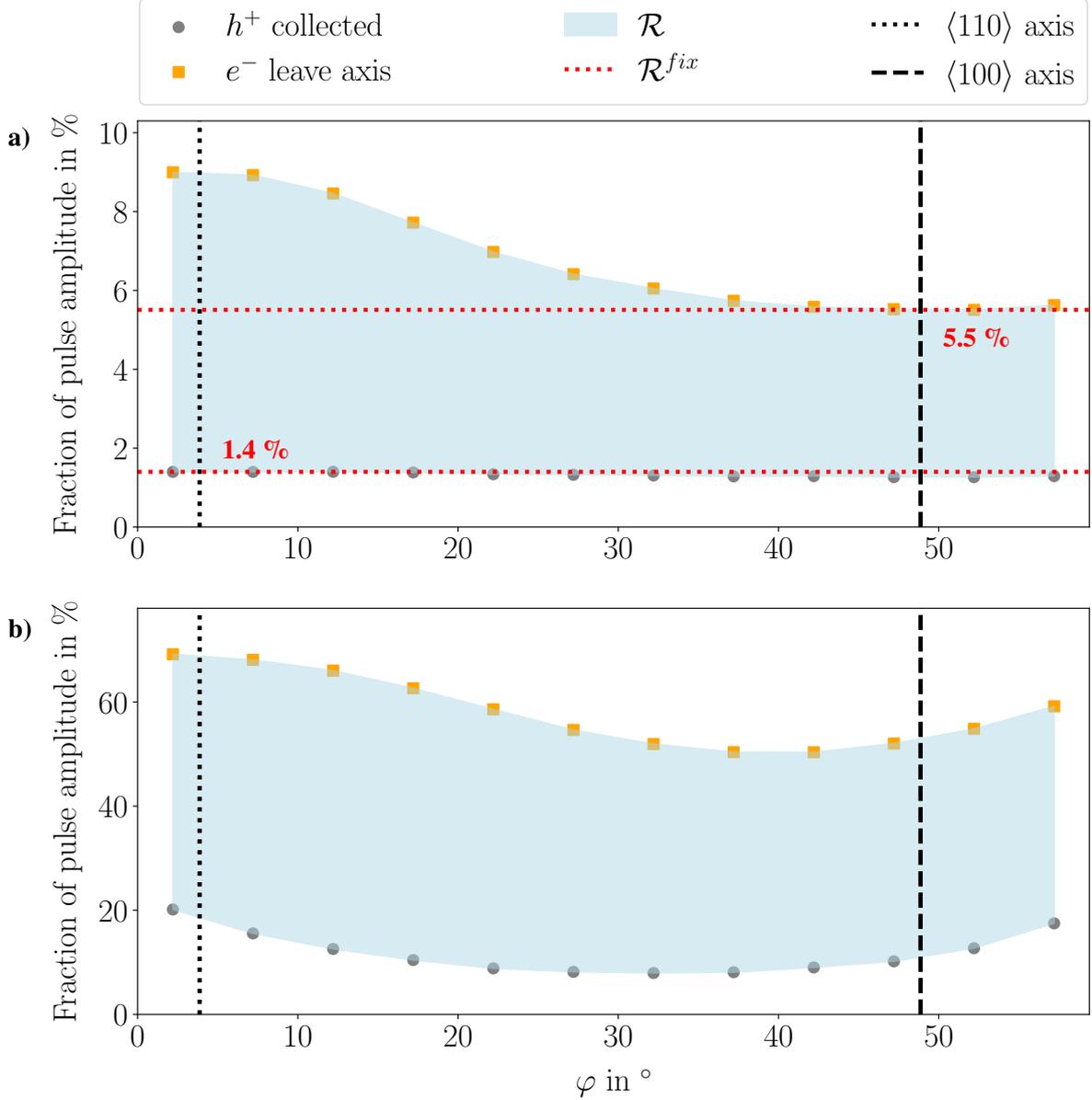
\begin{figure}[htb]
\centering
\begin{overpic}[width = \textwidth, , tics = 10]{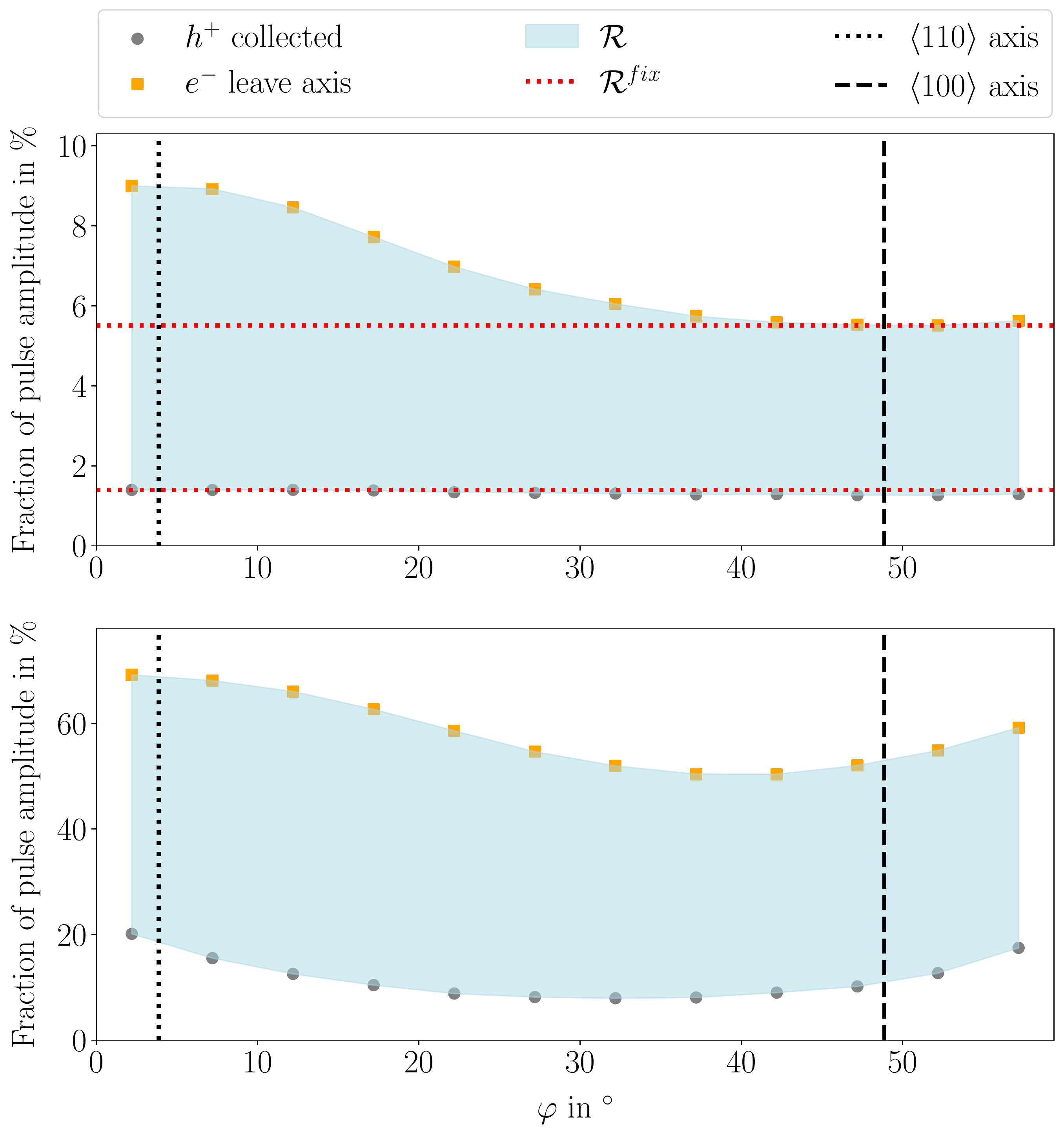}
\put (-3, 86) {\textbf{a)}}
\put (-3, 42) {\textbf{b)}}

\put (16,58) {\color{red} \textbf{1.4 \%}}
\put (80,68) {\color{red} \textbf{5.5 \%}}
\end{overpic}
\caption{$\mathcal{R}$ as obtained from simulated a)~Core and b)~Seg.\,1 pulses after the application of the respective response function for the $\varphi$ range from 2.2$^{\circ}$ to 57.2$^{\circ}$ at $r = 35.0$\,mm and $z=20.2$\,mm. The dotted red lines in a) indicate the choice of a fixed rise-time window.}
\label{fig:crtws_core_and_seg1_81keV}
\end{figure}

Custom rise-time windows were determined for all $\varphi$ for which data were taken in AS-1. These $\mathcal{R}$, both for the Core and the collecting Seg.\,1 are shown in Fig.\,\ref{fig:crtws_core_and_seg1_81keV}.\\

The electric field and the Core weighting potential are symmetric in $\varphi$, but vary in $r$ and $z$. The $\mathcal{R}$ are nevertheless $\varphi$ dependent, as the position where the electrons turn and the point in time when the holes are collected vary with $\varphi$ due to the effects of the crystal axes. Therefore, the weighting potential values at $t_{hc}$ and especially $t_{et}$ are different. In addition, the electric-field strengths along the different paths between  $t_{hc}$ and $t_{et}$ are different. Thus, in order to compare drift times at different $\varphi$, a fixed rise-time window, $\mathcal{R}^{fix} = 1.4\,\%-5.5\,\%$, which is contained in all $\mathcal{R}$, was chosen to ensure equivalent drift paths and, hence, the same electric field for evaluation in all $\varphi$.\\

\begin{figure}[htb]
    \centering
    \begin{overpic}[width = .9\textwidth,,tics = 10]{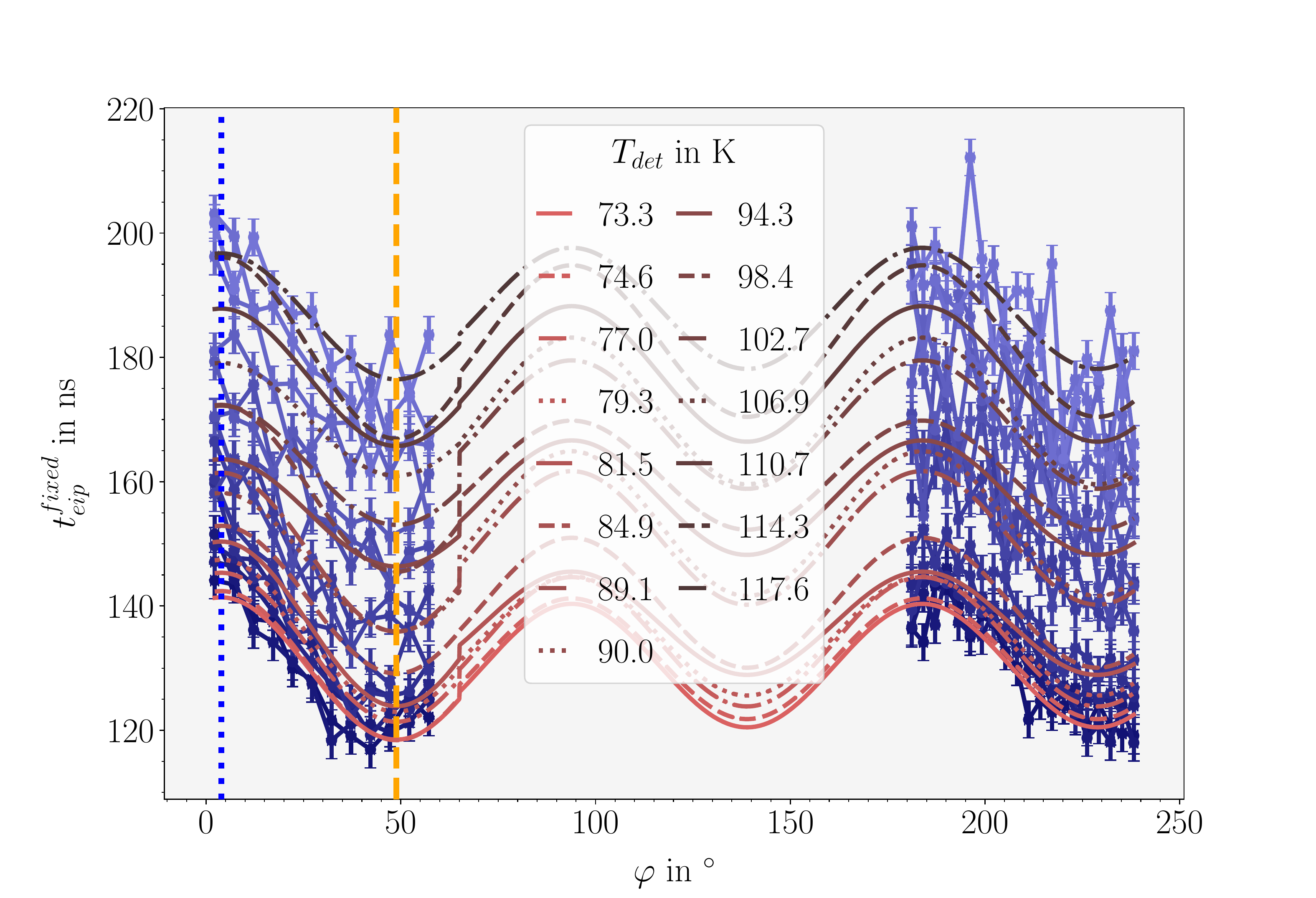}
    
    \end{overpic}
    \caption{Core rise-times, as determined for the fixed custom rise-time window, of 81\,keV Core superpulses versus $\varphi$ together with fits according to Eq.\,\eqref{eq:modified_sine}. The dotted\,(dashed) line indicates the location of the $\langle110\rangle$\,($\langle100\rangle$) axis. The vertical error bars represent the 3\,ns uncertainty on $t_{eip}^{fixed}$.}
    \label{fig:t_rt_fixed_over_phi_w_fit}
\end{figure}

\begin{figure}[htb]
    \centering
    \begin{overpic}[width = .9\textwidth,, tics = 10]{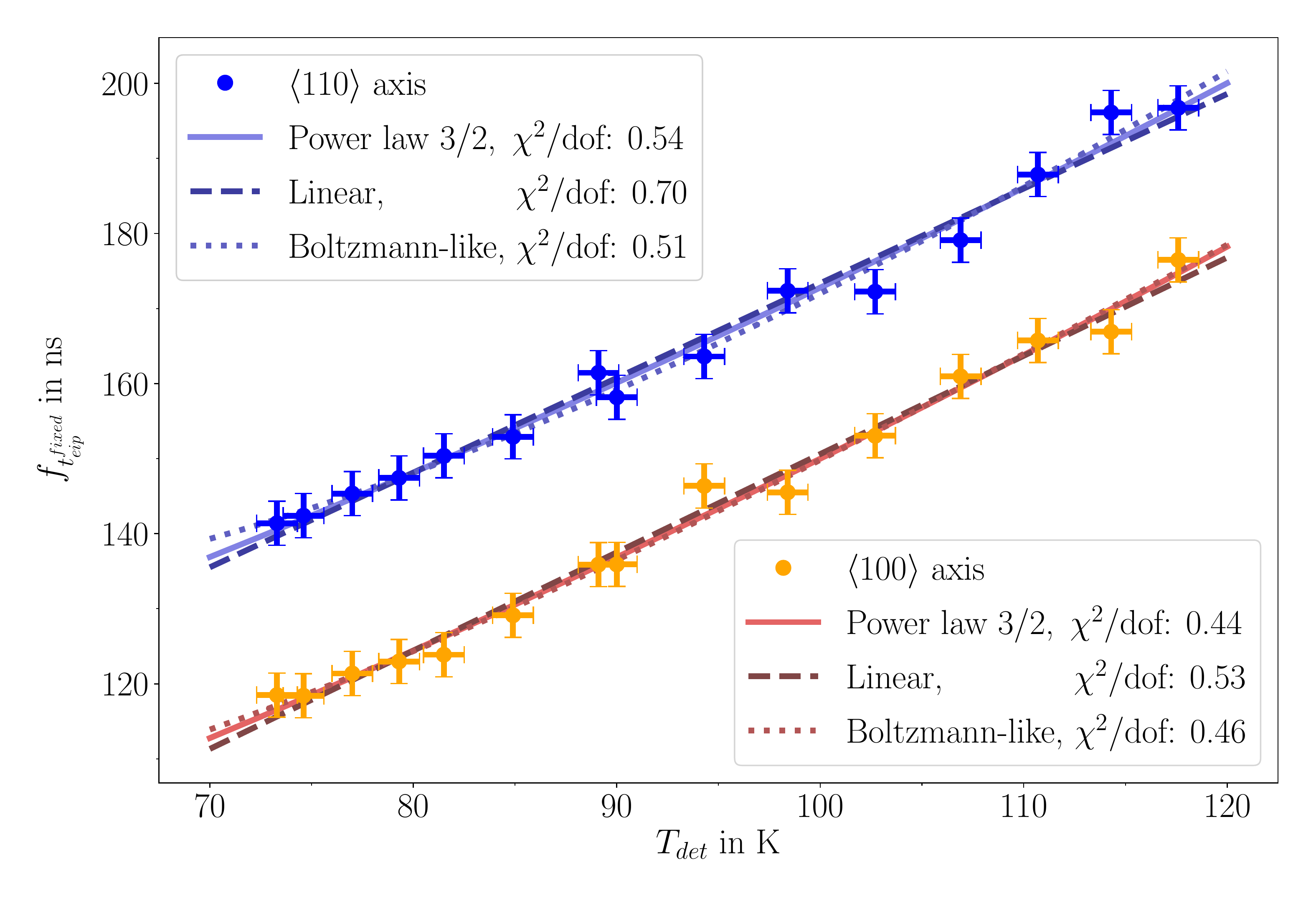}
    \put(55, 60) { \textbf{Core}}
    \put(21,18) {\vector(0,1){8}}
    \put(21,26) {\vector(0,-1){8}}
    \put(23,22.5) {\textbf{21.2\,\%}}
    
    \put(90,51) {\vector(0,1){7}}
    \put(90,58) {\vector(0,-1){7}}
    \put(80,51) {\textbf{13.1\,\%}}
    \end{overpic}
    \caption{Core rise-times, $f_{t^{fixed}_{eip}}$, as determined from the fits, see Eq.\,\eqref{eq:modified_sine} and Fig.\,\ref{fig:t_rt_fixed_over_phi_w_fit}, for both major axes, i.e.~at $\varphi = 48.87^{\circ}$ for $\langle$100$\rangle$ and $\varphi = 3.87^{\circ}$ for $\langle$110$\rangle$, versus $T_{det}$. Also shown are fits according to Eqs.\,\eqref{eq:powerlaw32}--\eqref{eq:boltzmannlike}.}
    \label{fig:t_rt_fixed_over_T_crtw}\label{fig:n-type_core_rtw_rise_times_versus_T}
\end{figure}

$\mathcal{R}^{fix}$ corresponds to a time interval for which electrons drift inwards and holes are already collected for all $\varphi$. The corresponding rise times, $t^{fixed}_{eip}$, as determined from Core superpulses are shown in Fig.\,\ref{fig:t_rt_fixed_over_phi_w_fit}, together with fits according to Eq.\,\eqref{eq:modified_sine}. The general trend of a decreasing anisotropy of drift velocities with increasing detector temperature is again observed as for $t_{rt}^{5-95}$. This is also seen in Fig.\,\ref{fig:n-type_core_rtw_rise_times_versus_T} which shows the evaluation of the fits, see Fig.\,\ref{fig:t_rt_fixed_over_phi_w_fit}, at the position of the crystallographic axes, in analogy to Fig.\,\ref{fig:risetimes_over_T_5-95}. The reduction of anisotropy is slightly less pronounced for the isolated axes than for the $t_{rt}^{5-95}$ rise times. Again, the three model functions, Eqs.\,\eqref{eq:powerlaw32}--\eqref{eq:boltzmannlike}, were fitted to the data. All three describe the data well. Table~\ref{tab:n-type_core_t_rt_vs_T_results} shows the corresponding fit parameters. It has to be noted that $\mathcal{R}^{fix}$ as determined from simulations at 78\,K was applied to measured superpulses for all temperatures.\\

\begin{figure}[htb]
    \centering
    \begin{overpic}[width = .9\textwidth, , tics = 10]{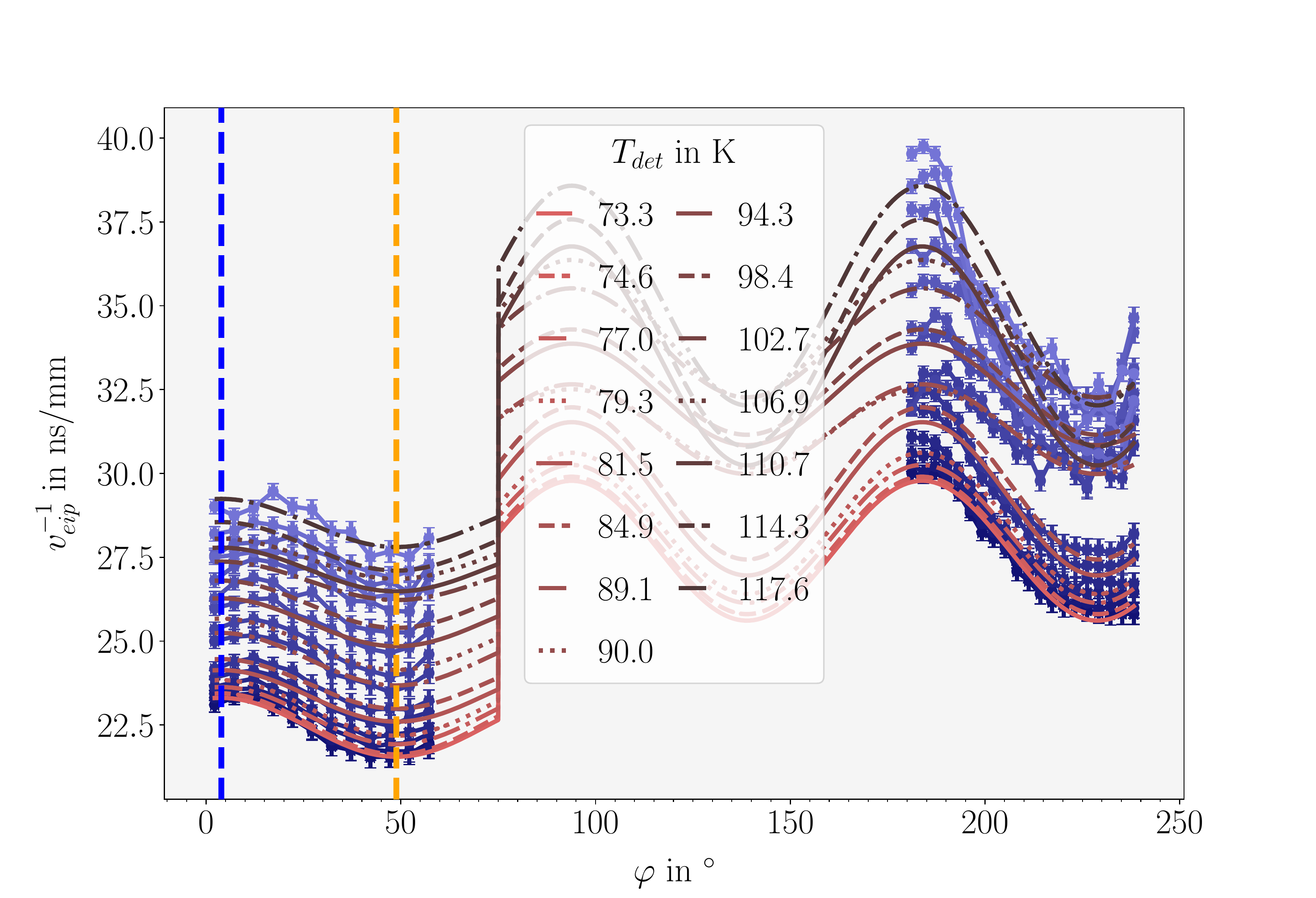}
    \put(20,35) {\textbf{Seg.\,1}}
    \put(75,16) {\textbf{Seg.\,4}}
    \end{overpic}
    \caption{Effective inverse velocities, $v_{eip}^{-1}$, as determined from 81\,keV Seg.\,1 and Seg.\,4 superpulses versus $\varphi$ together with fits according to Eq.\,\eqref{eq:modified_sine}. The dotted\,(dashed) line indicates the location of the $\langle110\rangle$ $\langle100\rangle$) axis. The vertical error bars represent the uncertainty on $v_{eip}^{-1}$, which was estimated to be
3\,ns divided by the individual drift-path lengths.}\label{fig:seg1_v_inv_over_phi_w_fit}
\end{figure}

\begin{figure}[htb]
    \centering
    \begin{overpic}[width = .9\textwidth,, tics = 10]{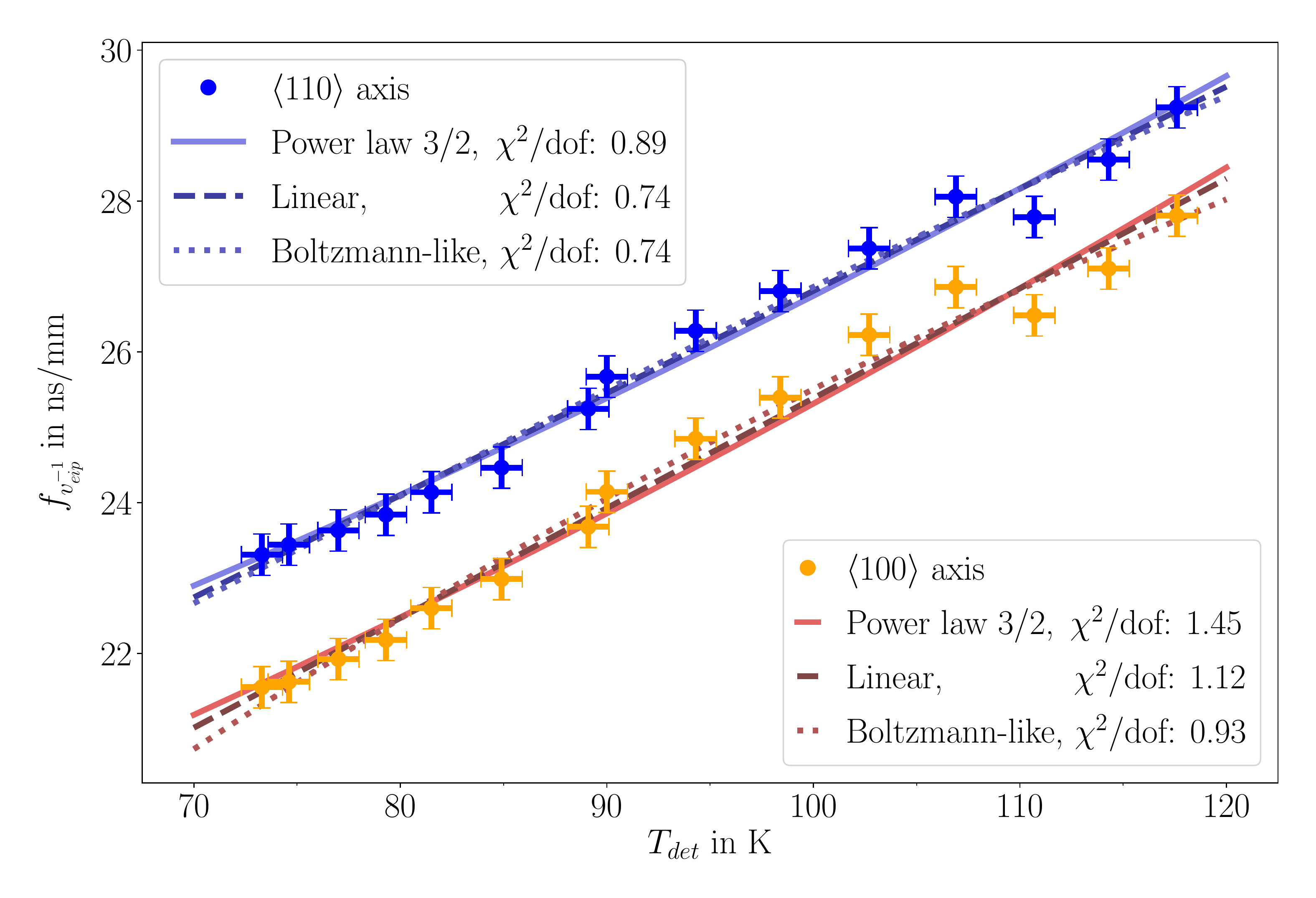}
    \put(54, 60) { \textbf{Seg.\,1}}
    \put(19,18) {\vector(0,1){6}}
    \put(19,24) {\vector(0,-1){6}}
    \put(21,22.5) {\textbf{8.1\,\%}}
    
    \put(91,55) {\vector(0,1){4}}
    \put(91,59) {\vector(0,-1){4}}
    \put(88,47) {\textbf{5.2\,\%}}
    \end{overpic}
    \caption{Effective inverse velocities, $f_{v^{-1}_{eip}}$, as determined from the fits, see Eq.\,\eqref{eq:modified_sine} and Fig.\,\ref{fig:seg1_v_inv_over_phi_w_fit}, for both major axes in Seg.\,1, i.e.~$\varphi = 48.87^{\circ}$ for $\langle$100$\rangle$ and $\varphi = 3.87^{\circ}$ for $\langle$110$\rangle$, versus $T_{det}$. Also shown are fits according to Eqs.\,\eqref{eq:powerlaw32}--\eqref{eq:boltzmannlike}.}
    \label{fig:n-type_seg1_rtw_rise_times_versus_T}
\end{figure}

For the segments, the weighting potentials are very different for different $\varphi$. Hence, a fixed rise-time window would correspond to largely different drift paths and the same analysis cannot be performed using segment data. Instead, the individual drift path can be used to define an effective inverse velocity\footnote{The velocity is called effective, since the rise time does not directly correspond to the real time it takes the electrons to travel the corresponding drift path, but also contains effects of the read-out electronics.}, $v_{eip}^{-1}$, as rise time over drift path length. For the $v_{eip}^{-1}$ determined from superpulses of Seg.\,1 and Seg.\,4 of AS-1 and AS-4, respectively, a sizable offset is observed, see Fig.\,\ref{fig:seg1_v_inv_over_phi_w_fit}. This is due to the significantly different response functions, see App.\,\ref{sec:response_functions}. As before, Eq.\,\eqref{eq:modified_sine} was fitted to the data points. The resulting values of $f_{v_{eip}^{-1}}$ at the location of the crystallographic axes in Seg.\,1 are shown for all $T_{det}$ in Fig.\,\ref{fig:n-type_seg1_rtw_rise_times_versus_T}.
Here, the change in anisotropy with increasing $T_{det}$ is significantly less pronounced than for the Core. It has to be noted that, while the $v^{-1}_{eip}$ are normalized by the respective drift path lengths, the field strengths are different for the  different drift paths. If the values of $\mathcal{E}$ were high enough to saturate the drift velocities, these differences would not matter and the anisotropy would be expected to be of the same order as for Fig.\,\ref{fig:n-type_core_rtw_rise_times_versus_T}. The observed difference in terms of longitudinal anisotropy as compared to Fig.\,\ref{fig:n-type_core_rtw_rise_times_versus_T} therefore shows that the electric-field strengths present here are well below the saturation region.\\

In general, the analysis presented here depends strongly on the similarity
of the shapes of the measured and the simulated pulses, as this is the prerequisite for meaningful applicability of the simulation based $\mathcal{R}$ to the data.
The simulation does not contain differential cross-talk, which is not corrected for in data. In addition, the knowledge on the impurity density profile is limited. However, the basic properties of the simulation are good enough for tests where the influence of these effects remains the same. This is the case for the temperature dependence. Thus, the influence of the employed charge drift models and their parameters can be tested.\\

The response function for the Core is the narrowest, see App.\,\ref{sec:response_functions}, and the Core pulses are the least affected
by cross-talk. In addition, the fixed rise-time window ensures very similar drift path and electric-field strengths for the two crystallographic axes. Therefore, the parameters obtained from applying the simulated $\mathcal{R}^{fix}$ to the measured superpulses for the Core, see Tab.\,\ref{tab:n-type_core_t_rt_vs_T_results}, are used for a further investigation of the implication of the measured $T$ dependence on the drift model.

\enlargethispage{4\baselineskip}

\begin{table}[htb]
\renewcommand*{\arraystretch}{1.2} 
    \centering
    \caption{Results of the fits according to Eqs.\,\eqref{eq:powerlaw32}--\eqref{eq:boltzmannlike} to the data shown in Fig.\,\ref{fig:n-type_core_rtw_rise_times_versus_T}. The experimental uncertainty refers to the 90\% confidence interval of the respective fit parameter as determined with \lsq{}.
    The systematic uncertainties result from fits to data including shifts of $\pm$\,1\,K. Where no systematic uncertainty is given, it was $<0.01$.}\label{tab:n-type_core_t_rt_vs_T_results}
    \vspace{0.2cm}
    \resizebox{\textwidth}{!}{%
   \begin{tabular}{l|c|ccc|r}
       & \textbf{Axis} & $\bm{p_0}$\,$\bm{\pm}$\textbf{(exp)}\,$\bm{^{+}_{-}}$\textbf{(syst)} & $\bm{p_1}$\,$\bm{\pm}$\textbf{(exp)}\,$\bm{^{+}_{-}}$\textbf{(syst)} & $\bm{p_2}$\,$\bm{\pm}$\textbf{(exp)}\,$\bm{^{+}_{-}}$\textbf{(syst)} & $\bm{\chi^2}$\textbf{/dof} \\
       \hline
       \multirow{2}{*}{\textbf{Eq.}\,\eqref{eq:powerlaw32}} & $\langle100\rangle$ & 60.2\,$\pm6.0$\,$^{+0.9}_{-0.9}$ & 0.09\,$\pm0.01$ & / & 0.44\\
       & $\langle110\rangle$ & 86.2\,$\pm6.0$\,$^{+0.8}_{-0.8}$& 0.09\,$\pm0.01$ & /& 0.54\\
       \hline
     \multirow{2}{*}{\textbf{Eq.}\,\eqref{eq:linear}} & $\langle100\rangle$ & 19.67\,$\pm8.8$\,$^{+1.3}_{-1.3}$ & 1.31\,$\pm0.09$ & / & 0.53\\
       & $\langle110\rangle$ & 47.2$\pm8.8$\,$^{+1.3}_{-1.3}$ & 1.26\,$\pm0.09$ & / & 0.70 \\
       \hline
       \multirow{2}{*}{\textbf{Eq.}\,\eqref{eq:boltzmannlike}} & $\langle100\rangle$ & 95.9\,$\pm19.0$\,$^{+0.3}_{-0.3}$ & 695\,$\pm515$\,$^{+11}_{-11}$ & 256\,$\pm111.0$\,$^{+3.5}_{-3.5}$ & 0.46\\
         & $\langle110\rangle$ & 127.1\,$\pm13.8$\,$^{+0.2}_{-0.2}$ & 933\,$\pm790$\,$^{+19}_{-18}$ & 303.3\,$\pm117.6$\,$^{+4.6}_{-4.5}$ & 0.51\\
         \hline
         \hline
     \end{tabular}
     }
      \vspace{0.4cm}
\end{table}%


\FloatBarrier
\section{Modeling Temperature Dependence of the Electron Drift}
The Boltzmann-like model, i.e.~Eq.\,\eqref{eq:boltzmannlike}, with parameters as determined from Core data using the fixed custom rise-time window, see Tab.\,\ref{tab:n-type_core_t_rt_vs_T_results}, was implemented to model the temperature dependence of the electron mobility in the charge-drift simulation. For the holes, results from a different data set taken with an almost identical p-type detector were used\,\cite{PhD:Schuster2021}. The holes are not important for the further analysis as their contribution to the Core pulses is excluded by the rise-time windows.\\

\begin{figure}[htb]
    \centering
    \begin{overpic}
        [width = .9\textwidth, , tics = 10]{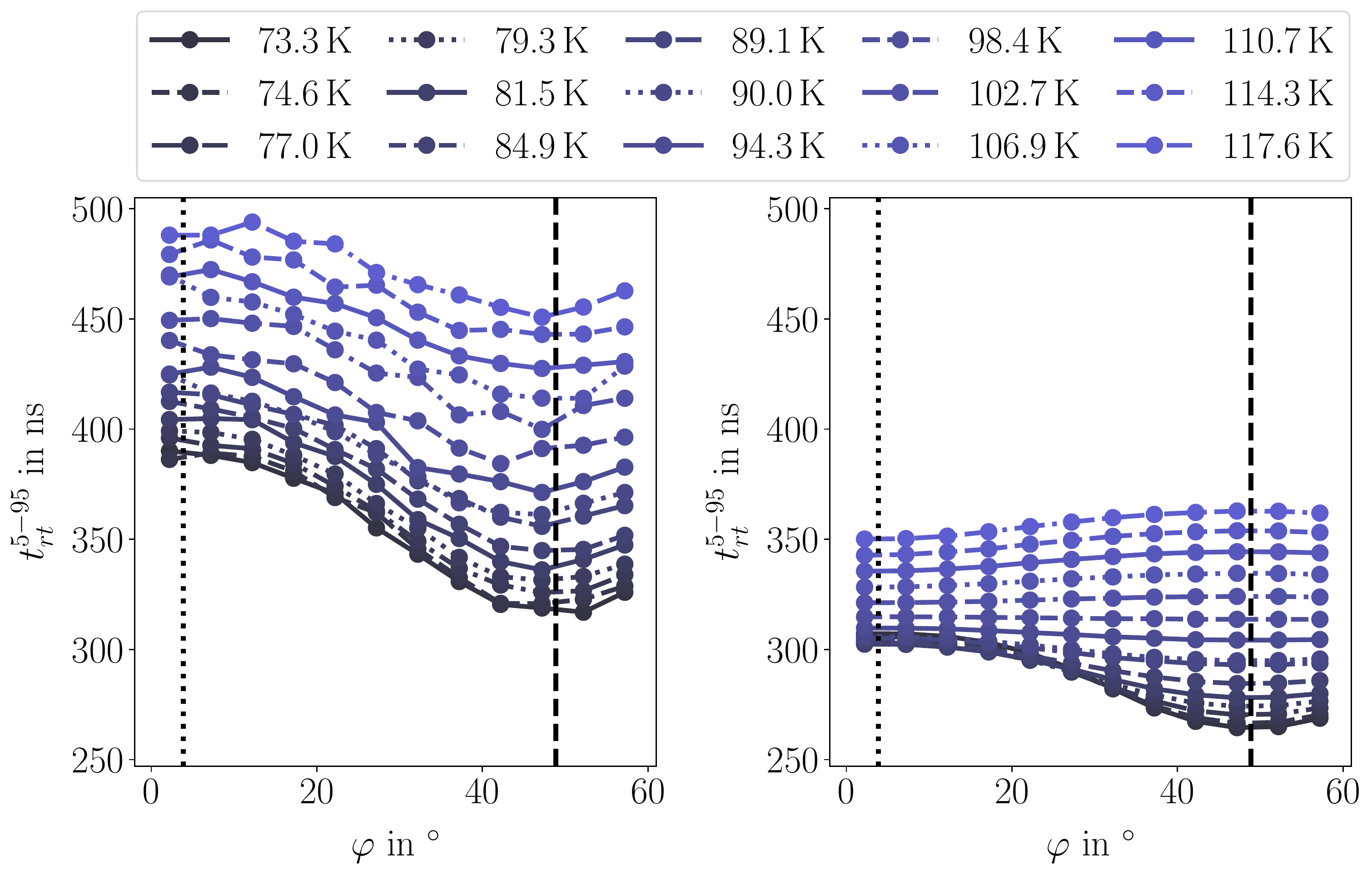}
        \put(0.7, 47.5) {\textbf{a)}}
        \put(50.5, 47.5) {\textbf{b)}}
        \put (15, 18) {\small\textbf{Data}}

        \put (65.5, 46) {\small\textbf{Simulation}}
        \put (65.5, 42) {\small\textbf{Default charge-}}
        \put (65.5, 39) {\small\textbf{drift models}}
        \put (65.5, 35) {\small\textbf{$+$ Boltzmann-like}} 
        \put (65.5, 32){\small\textbf{temperature model}}
    \end{overpic}
    \caption{Comparison of $t_{rt}^{5-95}$ as determined from Core superpulses\,(left) and from simulated Core pulses\,(right), using the default electron drift model and the Boltzmann-like temperature model. The dotted\,(dashed) black line marks the position of the $\langle110\rangle$\,($\langle100\rangle$) axis.}\label{fig:ntype_rt5_95_anisotropy_default_all_T}
\end{figure}

\begin{figure}[htb]
    \centering
    \begin{overpic}
        [width = .9\textwidth, ,tics = 10]{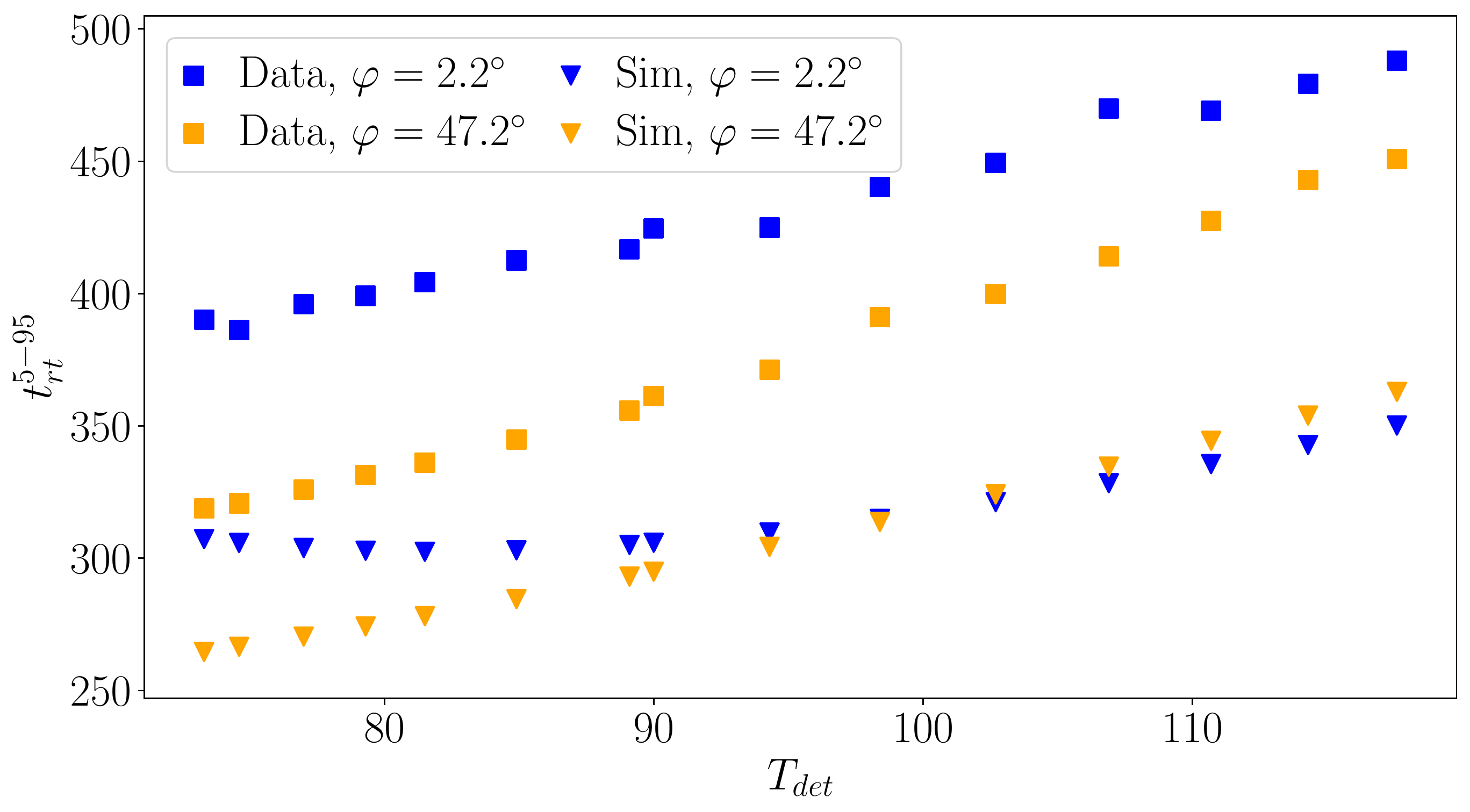}
    \end{overpic}
    \caption{Comparison of $t_{rt}^{5-95}$ versus $T_{det}$ at $\varphi = 2.2^{\circ}$ and $\varphi = 47.2^{\circ}$ between data, see Fig.\,\ref{fig:ntype_rt5_95_anisotropy_default_all_T}a, and simulation using the default electron-drift model and the Boltzmann-like temperature model, see Fig.\,\ref{fig:ntype_rt5_95_anisotropy_default_all_T}b.}
    \label{fig:t5-95_on_axes_data_vs_sim_all_T}
\end{figure}

The parameters for the $T$ dependence were measured for the $\langle100\rangle$ and $\langle110\rangle$ axes and not for the $\langle111\rangle$ axis. Thus, Eqs.\,\eqref{eq:ve110}~and~\eqref{eq:vh110} were first used to calculate $v_{e}^{110}$ and $v_{h}^{110}$ at the reference temperature of 78\,K. The results, together with $v_{e}^{100}$ and $v_{h}^{100}$, were then scaled according to the temperature model.\footnote{Generally, the temperature  dependence of $v$ is itself also dependent on $\mathcal{E}$ via the denominator terms in Eq.\,\eqref{eq:drift_velocity} containing the temperature-dependent factors $\beta$ and $\mathcal{E}_0$. It has been verified that for the electric-field strengths occurring during the fixed rise-time window, the influence of the electric field on the temperature dependence is negligible.} The scaled velocities at the respective temperature, $v_{e,\,T_{det}}^{110}$\,($v_{h,\,T_{det}}^{110}$), were then used together with the scaled $v_{e,\,T_{det}}^{100}$\,($v_{h,\,T_{det}}^{100}$) to calculate $v_{e,\,T_{det}}^{111}$\,($v_{h,\,T_{det}}^{111}$).\\
A direct comparison between the resulting predictions from simulation and AS-1 data for the $t_{rt}^{5-95}$ of the Core pulses is shown versus $\varphi$ in Fig.\,\ref{fig:ntype_rt5_95_anisotropy_default_all_T}. As already discussed in Sec.\,\ref{sec:Tdep-rt5-95}, the data show a decrease in anisotropy and general shift towards higher $t_{rt}^{5-95}$ with rising $T_{det}$. In the simulation, the overall level of $t_{rt}^{5-95}$ is lower than in the data. This could be explained by the level of the electrically active impurities being slightly off in the simulation and/or a different level of neutral impurities in the detector under study versus the detector for which the mobility parameters were measured. More important are the following problems with the simulation.\\
For the first few temperatures, the absolute value of $t_{rt}^{5-95}$ for events generated on the $\langle110\rangle$ axis is decreasing instead of increasing, see Fig.\,\ref{fig:t5-95_on_axes_data_vs_sim_all_T}. Furthermore, only for the lowest ten $T_{det}$, the simulated $t_{rt}^{5-95}$ is higher for the $\langle110\rangle$ axis than for the $\langle100\rangle$ axis. These features are clearly not observed in data.\\

\begin{figure}[htb]
\centering
      \begin{overpic}
          [width=.9\textwidth, , tics = 10]{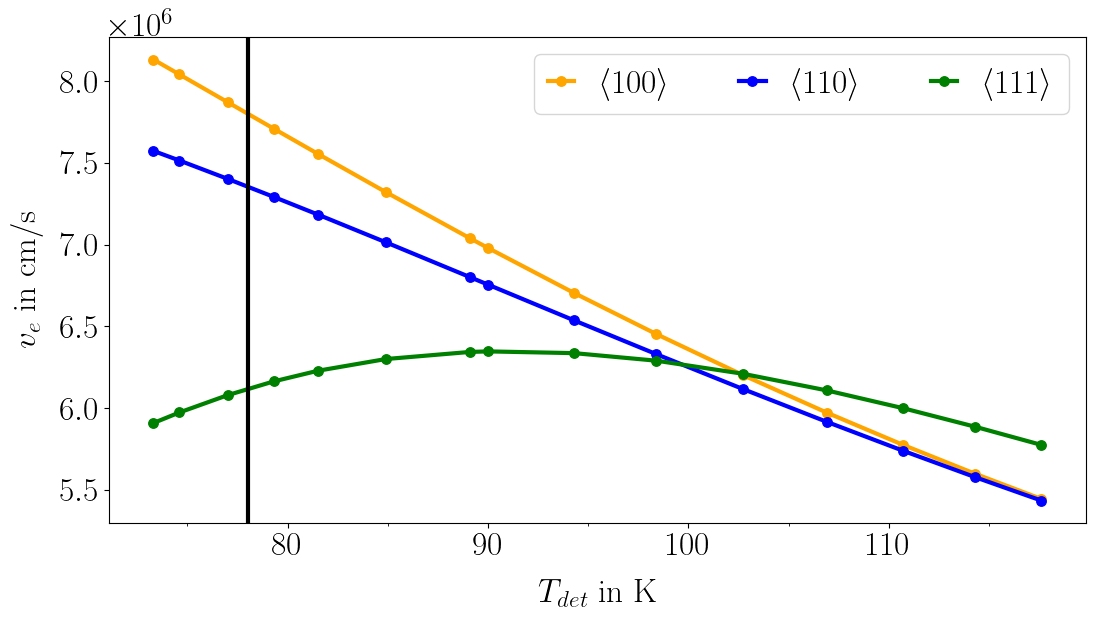}
          \put (26,14) {\textbf{Default electron drift model}}
          \put (26, 11) {\textbf{$+$ Boltzmann-like temperature model}} 
      \end{overpic}
    \caption{Electron drift velocities, $v_{e}^{100}$, $v_{e}^{110}$ and $v_{e}^{111}$ versus $T_{det}$ for an electric-field strength of 439\,V/cm as calculated using the default drift velocity model in combination with the Boltzmann-like temperature model, i.e.\,Eq.\,\eqref{eq:boltzmannlike} with the parameters of Tab.\,\ref{tab:n-type_core_t_rt_vs_T_results}.
    The black line indicates the reference temperature of 78\,K.
    }\label{fig:electron_velocities_vs_T}
\end{figure}

Figure~\ref{fig:electron_velocities_vs_T} shows calculated electron drift velocities for the three major axes versus $T_{det}$ for an electric-field strength of $\bar{\mathcal{E}}_{dp}\,=\,439$\,V/cm, which is the mean electric-field strength in the segBEGe along the drift paths\footnote{The individual $\mathcal{E}$ ranged from $\approx$\,410\,V/cm to $\approx$\,490\,V/cm.} corresponding to the $\mathcal{R}^{fix}$, see Sec.\,\ref{sec:CRTW}. Note, that this serves only as a visualization, as it is calculated for a specific $\mathcal{E}$. In the charge drift simulation, the velocities are calculated at each time step using the $\mathcal{E}$ present at the respective position.\\
The value of $v_e^{111}$ is important because a significant part of the drift is influenced by the \br{111} axis. The initial rise of $v_e^{111}$ with $T_{det}$ causes the prediction that $t_{rt}^{5-95}$ decreases with $T_{det}$ for the first five temperatures for events first drifting along the $\langle110\rangle$ axis. 
This is not observed in data and is unphysical.\\

Even at the reference temperature of 78\,K, independent of the implemented temperature model, data and simulation show a mismatch in terms of anisotropy of $t_{rt}^{5-95}$.
As shown in Fig.\,\ref{fig:ntype_rt5_95_anisotropy_default_all_T}, the anisotropy between the $\langle100\rangle$ and $\langle110\rangle$ axes at 77.0\,K, i.e.~the highest and lowest $t_{rt}^{5-95}$, is underestimated as $\Delta t_{rt}^{5-95}\,=\,29.6$\,ns in the simulation as compared to $\Delta t_{rt}^{5-95}\,=\,66.8$\,ns for data. In addition, the simulated rise times are overall shorter.
The absolute discrepancy between $t_{rt}^{5-95}$ in data and simulation for the $\langle100\rangle$ axis could be partially explained by the imperfect knowledge on the impurity density profile influencing the electric-field strengths within the detector, especially close to the surface. However, this would affect the drift along both axes in the same way\footnote{A $\varphi$-symmetry for the impurity density profile seems like a fair assumption for crystals pulled via the Czochralski method\,\cite{czochralski1918:CrystalGrowth}, like the one under study.} and, thus, cannot explain the differences in anisotropy.\\
In essence, the implementation of the temperature dependence using the Boltzmann-like temperature model leads to predictions contradicting the measurements from which its parameters were extracted. 
These results, therefore, strongly indicate that the default electron-drift model needs to be revisited.\\ 

The default electron drift velocity is given by a weighted sum of terms, $\frac{\gamma_j}{\sqrt{\bm{\mathcal{E}}_n(\bm{r})^\top\gamma_j\bm{\mathcal{E}}_n(\bm{r})}}$, which describe the projection of the reciprocal effective mass tensor, $\gamma_{j}$, along the normalized electric field, $\bm{\mathcal{E}}_{n}(\bm{r})$ see App.\,\ref{app:e_drift_model}, Eq.\,\eqref{eq:eledrifta}.\footnote{With $\gamma_j \propto (m_{e}^{*})^{-1}$.} These terms scale with
$(m_{e}^{*})^{-1/2}$. This dependence of the electron velocity on $m_{e}^{*}$ reflects an assumption that the ionized impurities are the predominant scattering centers in the crystal\,\cite{Mihailescu:2000jg,Conwell1950:ImpurityScattering}. In high-purity germanium, however, the concentration of ionized impurities is at the very low level of $\approx10^{10}$\,cm$^{-3}$. Therefore, this assumption seems questionable. Instead, it is fair to assume, that the scattering off acoustic phonons is the dominating process as already discussed in Sec.\,\ref{sec:tdep}. This implies $\mu_{e} \propto (m_{e}^{*})^{-5/2}$.\\
As a first test, the exponent of the expression $(\,\bm{\mathcal{E}}_n(\bm{r})^\top\gamma_j\bm{\mathcal{E}}_n(\bm{r})\,)^{-1/2}$, which represents an effective weighting factor to $\gamma_{j}$\,\cite{Nathan1963:Anisotropy}, was changed to $3/2$ in all terms of the drift model,
i.e.~$(\,\bm{\mathcal{E}}_n(\bm{r})^\top\gamma_j\bm{\mathcal{E}}_n(\bm{r})\,)^{-1/2}$ $\rightarrow (\,\bm{\mathcal{E}}_n(\bm{r})^\top\gamma_j\bm{\mathcal{E}}_n(\bm{r})\,)^{3/2}$, see App.\,\ref{app:e_drift_model}. This results in the overall expression $\gamma_j \cdot (\,\bm{\mathcal{E}}_n(\bm{r})^\top\gamma_j\bm{\mathcal{E}}_n(\bm{r})\,)^{3/2} \propto (m_{e}^{*})^{-5/2}$. Using this modified electron drift model, the parameters of Eq.\,\eqref{eq:ve110} change: $\beta^{new}_{100}\,=\,0.386$ and $\beta^{new}_{111}\,=\,0.690$.\\

\begin{figure}[htb]
    \centering
    \begin{overpic}
        [width = \textwidth, , tics = 10]{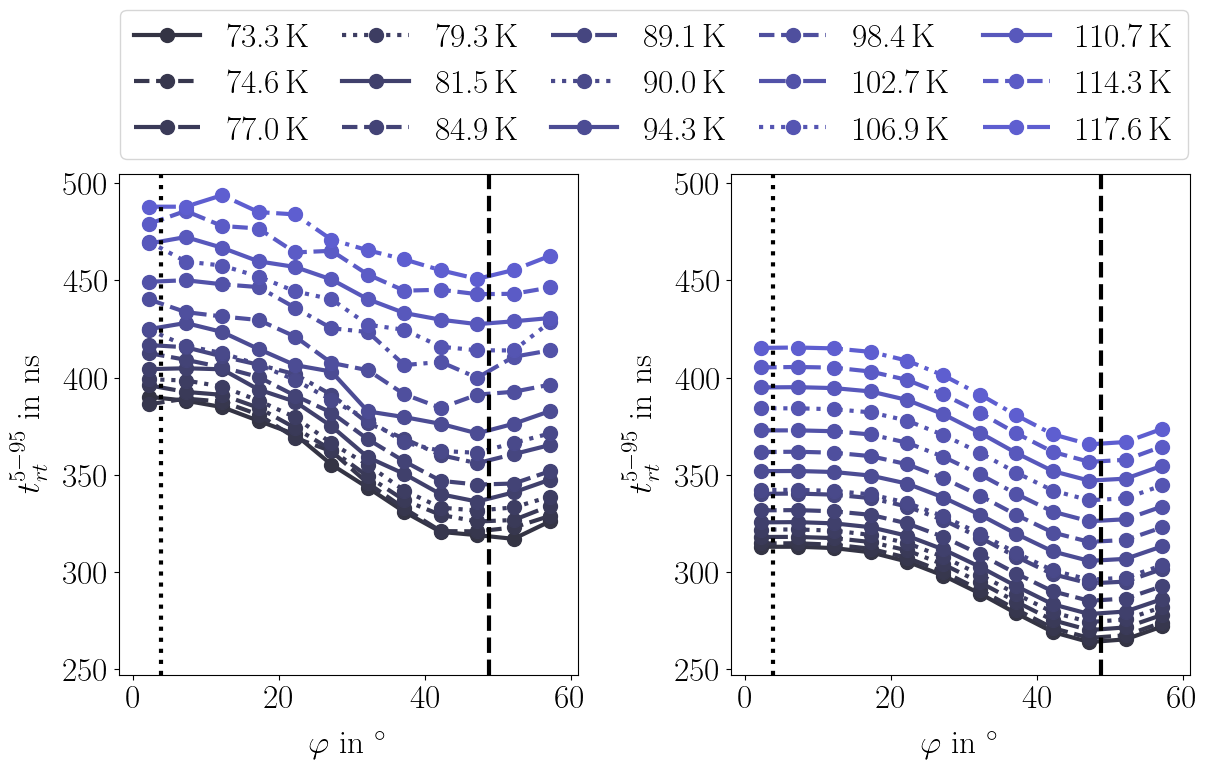}
        \put(0.7, 47.5) {\textbf{a)}}
        \put(50.5, 47.5) {\textbf{b)}}
        
        \put (15, 18) {\small\textbf{Data}}

        \put (65.5, 46) {\small\textbf{Simulation}}
        \put (65.5, 42) {\small\textbf{Modified electron}}
        \put (65.5, 39) {\small\textbf{drift model}}
        \put (65.5, 15) {\small\textbf{$+$}} 
        \put (65.5, 12) {\small\textbf{Boltzmann-like}} 
        \put (65.5, 9){\small\textbf{temperature model}}
    \end{overpic}
    \caption{Comparison of $t_{rt}^{5-95}$ as determined from Core superpulses\,(left) and from simulated Core pulses\,(right) using the modified electron drift model. The dotted\,(dashed) black line marks the position of the $\langle110\rangle$\,($\langle100\rangle$) axis.}
    \label{fig:ntype_rt5_95_anisotropy_newADL_all_T}
\end{figure}

\begin{figure}[htb]
\centering
  \begin{overpic}
      [width=\textwidth]{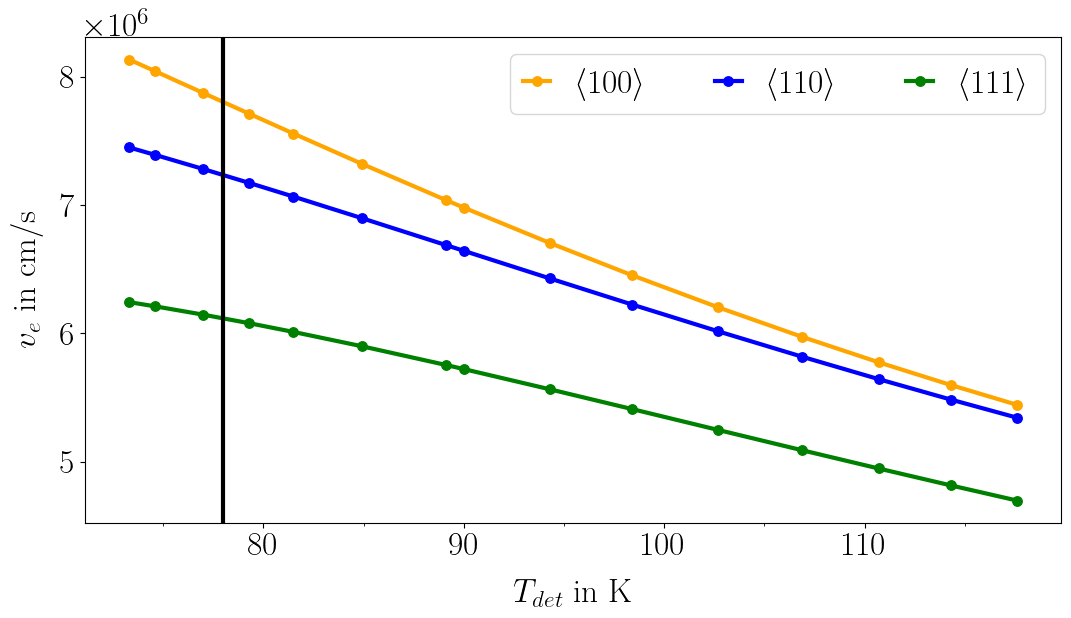}

      \put (26, 14) {\textbf{Modified electron drift model}}
       \put (26, 11) {\textbf{$+$ Boltzmann-like temperature model}} 
  \end{overpic}
  \caption{Electron drift velocities, $v_{e}^{100}$, $v_{e}^{110}$ and $v_{e}^{111}$ versus $T_{det}$ for an electric-field strength of 439\,V/cm using the modified electron drift model in combination with the Boltzmann-like temperature model, i.e.\,Eq.\,\eqref{eq:boltzmannlike} with the parameters of Tab.\,\ref{tab:n-type_core_t_rt_vs_T_results}. The black line indicates the reference temperature of 78 K.}\label{fig:electron_velocities_vs_T_ADL_new}
\end{figure}

Figure~\ref{fig:ntype_rt5_95_anisotropy_newADL_all_T} shows the comparison between predictions of the simulation based on the modified electron drift model together with the Boltzmann-like temperature model and data from AS-1 in terms of $t_{rt}^{5-95}$ of the Core pulses versus $\varphi$. The overall predicted level of $t_{rt}^{5-95}$ is higher than for the default electron-drift model and, while it is still lower than what is observed in data, the anisotropy at 77.0\,K of $\Delta t_{rt}^{5-95}\,=\,48.0$\,ns is closer to the value observed in data than the default prediction. Overall, the predictions of the temperature dependence from the simulation using the above described modifications to the electron drift model and the Boltzmann-like temperature model fit the data much better than the predictions based on the default electron drift model. The decline of anisotropy with $T_{det}$, however, is still not properly predicted.
Figure~\,\ref{fig:electron_velocities_vs_T_ADL_new} shows the corresponding calculated drift velocities over $T_{det}$ for the representative electric-field strength of 439\,V/cm in analogy to Fig.\,\ref{fig:electron_velocities_vs_T}.\\
It should be noted
that the parameters of the temperature model are based on the fixed custom rise-time window which was determined using the default electron drift model, see Sec.\,\ref{sec:CRTW}. Future iterations of this analysis will also include simulations using the modified electron drift model. The described modifications lead to a longer horizontal inwards drift for electrons at the middle height of the detector and the fixed rise-time window used here would still be contained and, thus, valid. Such iterative studies, together with improvements on the knowledge of the impurity density profiles from other analyses, could further refine the electron-drift model and the temperature-dependence model for the charge carrier drift.\\

\begin{figure}[htb]
    \centering
    \begin{overpic}
        [width = .9\textwidth, ,tics = 10]{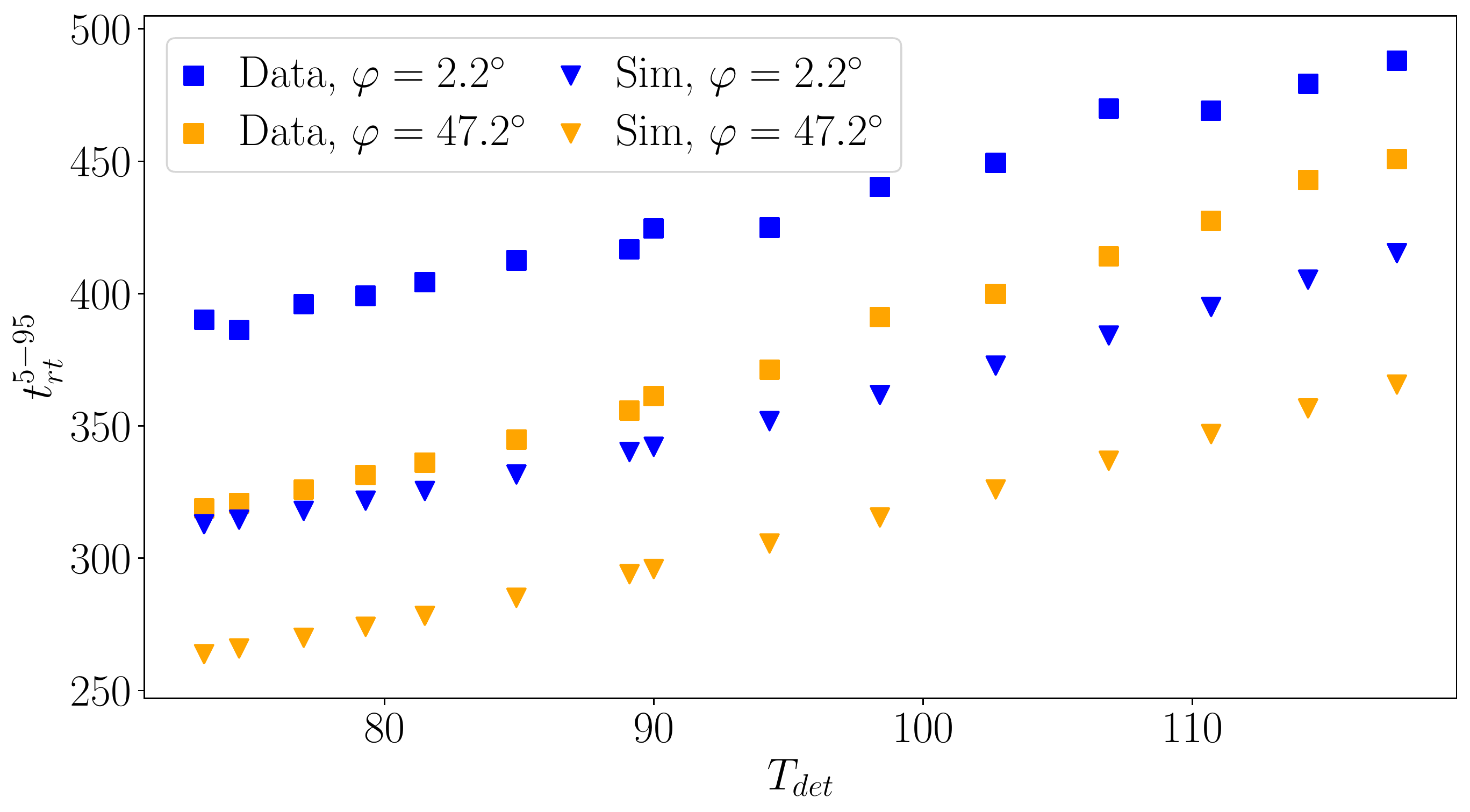}
    \end{overpic}
    \caption{Comparison of $t_{rt}^{5-95}$ versus $T_{det}$ at $\varphi = 2.2^{\circ}$ and $\varphi = 47.2^{\circ}$ between data, see Fig.\,\ref{fig:ntype_rt5_95_anisotropy_newADL_all_T}a, and simulation using the modified electron-drift model and the Boltzmann-like temperature model, see Fig.\,\ref{fig:ntype_rt5_95_anisotropy_newADL_all_T}b.}
    \label{fig:t5-95_on_axes_data_vs_sim_all_T_newADL}
\end{figure}

\FloatBarrier
\section{Summary and Outlook}
Data were taken and analyzed for an n-type four-fold segmented point-contact detector at different temperatures between 73\,K and 118\,K. The dependence of the rise time of the pulses on the detector temperature was investigated. It was shown that the rise time for pulses increases with higher temperatures while the anisotropy in rise time with respect to the different crystallographic axes decreases.\\
Using the pulse-shape simulation package \SSD{}, custom rise-time windows were derived from simulated drift paths and pulses at representative positions in an attempt to disentangle the temperature dependence for individual crystal axes. This can be seen as a first step in an iterative process. A fit of a Boltzmann-like function to the measured rise times yielded reasonable results.\\
This temperature model was implemented into \SSD{} together with the determined parameters as an attempt to reproduce the observations. The resulting simulations yielded nonphysical results. This was mainly attributed to inaccuracies of the electron drift model. A first attempt to modify the electron drift model by changing the assumptions on the dominant scattering centers yielded significantly improved results. This strongly suggests that the electron mobilities are predominantly limited by the scattering off acoustic phonons.\\
Further measurements could be taken with a coaxial detector. The much simpler drift paths in the plane spanned by the \br{100} and \br{110} axes would allow a more precise determination of the temperature dependence of the charge carrier drift using the presented method. In addition, information on the density distribution of electrically active impurities in a test detector provided by complementary measurements\,\cite{Bruyneel2011:space_charges,Birkenbach2011:charge_distribution,Hauertmann2022:Impurities} would improve the scientific rigor of this method and also open the possibility to put limits on neutral impurities. The electron-drift model itself could then be adjusted further. It is probably necessary to consider not only one dominating scattering process but a mixture of processes.
\clearpage
\renewcommand{\thesection}{\Alph{section}}
\renewcommand{\thesubsection}{\Alph{section}\,-\,\Roman{subsection}}
\renewcommand{\thefigure}{\Alph{section}\,-\,\arabic{figure}}
\setcounter{section}{0}
\appendix
\counterwithin{figure}{section}
\counterwithin{table}{section}
\section{Crystal Structure of Germanium}\label{app:germanium_properties}
Germanium is an element of group \RNum{4} of the periodic table with the atomic number $Z=32$. It forms a face-centered cubic, $fcc$, crystal structure like diamond with a two-atom base.

\begin{figure}[htb]
\begin{minipage}{0.4\textwidth}
    \centering
    \captionof{table}{Points of interest and sets of directions of high symmetry for the first Brillouin zone of an \mbox{$fcc$ lattice}.}
    \begin{tabular}{={11} ={-1}}
        \mc{1}{l}{\tbf{Directions}} & \mc{1}{r}{\tbf{Points}}\\
        \hhline{==}\\
        \mc{1}{l}{In Miller indices} & \Gamma = (000)\\
        \Delta = \langle 100\rangle & \text{X} = (010)\\
        \Sigma = \langle 110\rangle & \text{K} = (110)\\
        \Lambda = \langle 111\rangle & \text{L} = (111)\\
        \hline
    \end{tabular}
    \label{tab:brillouin}
    \vspace{.3cm}
    \begin{overpic}[width=\textwidth,,tics=10]
        {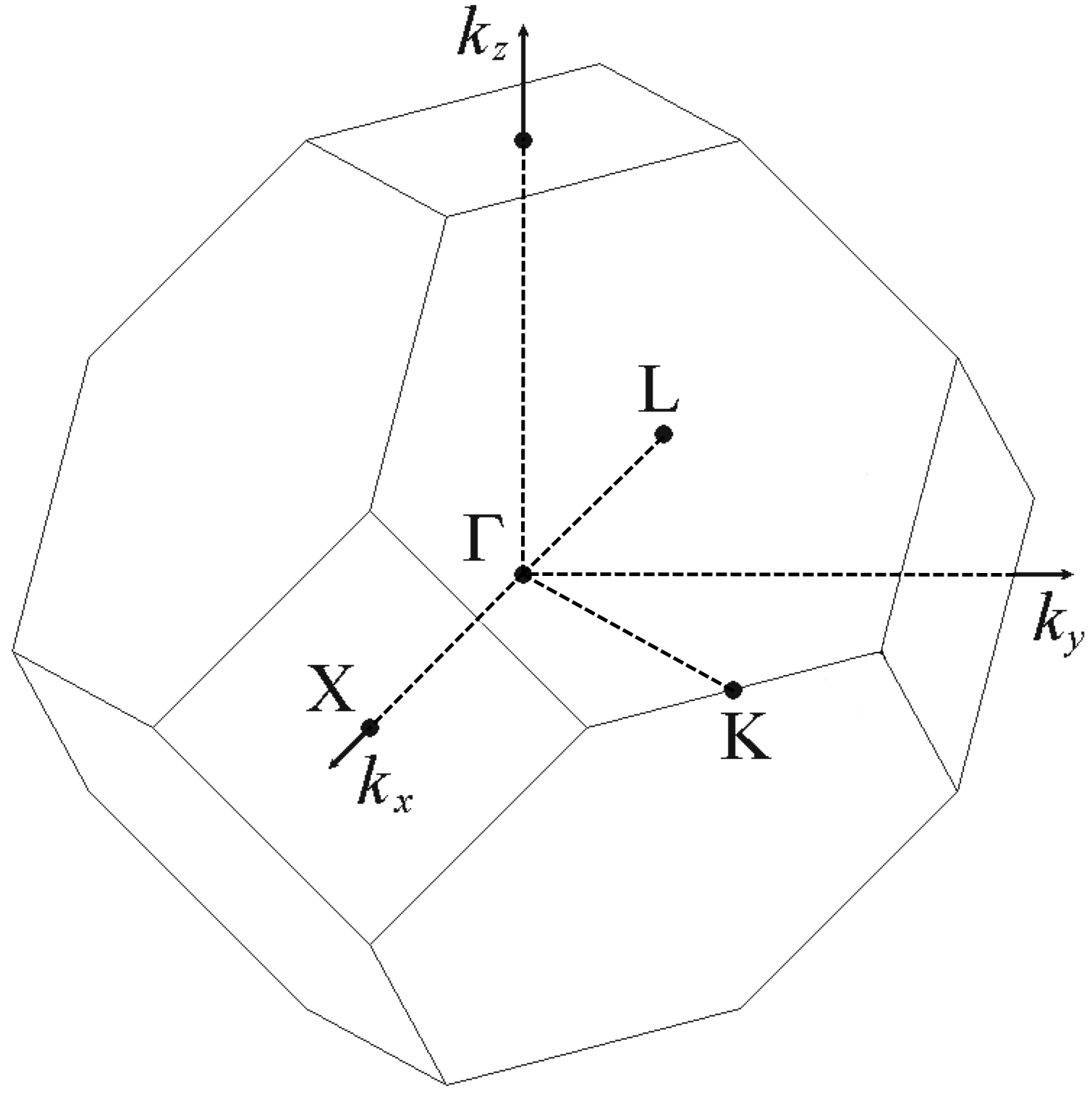}
        \put(40,36){$\Delta$}
        \put(53,38){$\Sigma$}
        \put(51,55.5){$\Lambda$}
    \end{overpic}
    \captionof{figure}{First Brillouin zone of an \mbox{$fcc$ lattice}.}
    \label{fig:fcc_brillouin}
        \vspace{1cm}
\end{minipage}
\hfill
\begin{minipage}{0.55\textwidth}
    \centering
    \begin{overpic}[width = \textwidth,,tics=10]
        {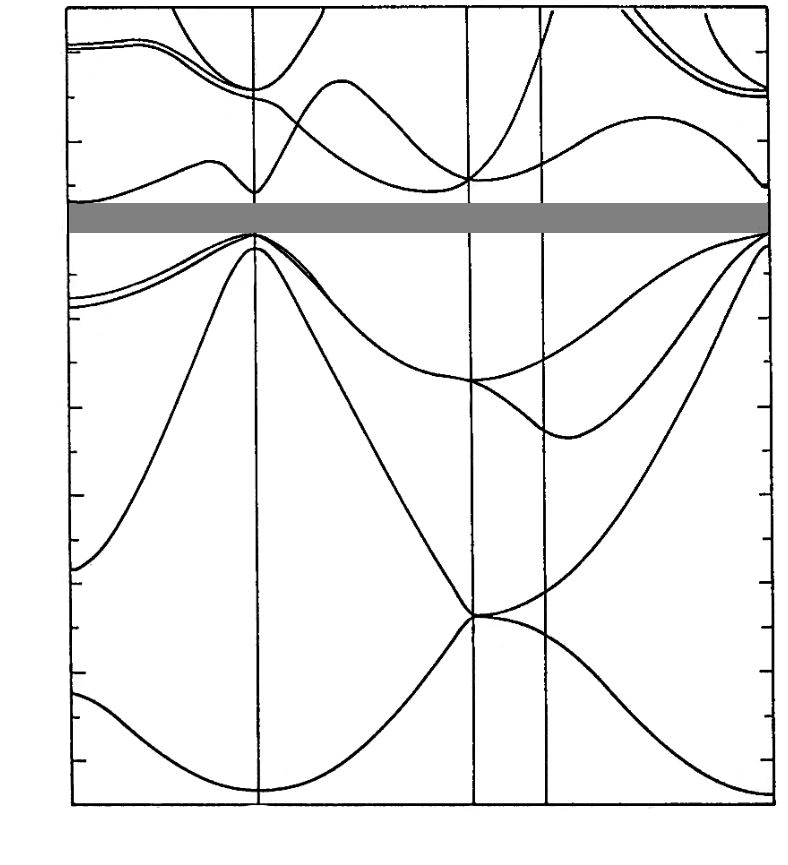}
        \put(8,3){L}
        \put(19,3){$\Lambda$}
        \put(29,3){$\Gamma$}
        \put(41,3){$\Delta$}
        \put(53.5,3){X}
        \put(62,3){K}
        \put(75,3){$\Sigma$}
        \put(89,3){$\Gamma$}
        \put(42, 74){\textbf{\footnotesize{band gap}}}
        \put(1,11){-12}
        \put(1,21){-10}
        \put(3,32){-8}
        \put(3,42){-6}
        \put(3,52){-4}
        \put(3,62){-2}
        \put(4,72){0}
        \put(4,82){2}
        \put(4,93){4}
        \put(0,42){\turnbox{90}{Energy in eV}}
        \put(38,-2){Wave vector $\mathbf{k}$}
    \end{overpic}
    \caption[The band structure of germanium.]{Band structure of germanium. Energy bands along specific $\mathbf{k}$ vectors are emphasized by vertical lines, see Fig.\,\ref{fig:fcc_brillouin}.
    The band gap is indicated as a shaded region. The maximum of the valence band is set to 0\,eV. In this figure, moving along the x-axis corresponds to changing the $\mathbf{k}$ vector on a straight line between the labeled points in the Brillouin zone. Figure adapted from~\cite{Cohen1988:GeBandStructure}.}
    \label{fig:Ge_Bandstructure}
\end{minipage}
\end{figure}
The Brillouin zone is shown in Fig.\,\ref{fig:fcc_brillouin}.    Germanium crystals have three major crystallographic axes, $\langle 100\rangle$, $\langle 110\rangle$ and $\langle 111\rangle$, where the lattice is invariant under rotations.\\
The detailed band structure of germanium is shown in Fig.\,\ref{fig:Ge_Bandstructure}. The minima and maxima in the band structure, especially the ones closest to the band gap can be locally approximated by parabola. This way, the energy of the electrons and holes can be described in analogy to the energy of free electrons using an effective mass, $m^{*}$, which depends on the local curvature. Charge carriers in an electric field attempt to drift along the electric field lines. At the L-point, i.e.~the conduction band minimum in germanium, see Tab.\,\ref{tab:brillouin} and Figs.\,\ref{fig:fcc_brillouin}~and~\ref{fig:Ge_Bandstructure}, the movement along $\Lambda$ is called longitudinal and the resulting effective electron mass is $m^*_{e,L} = 1.64 \cdot m_e$. Along a transverse direction, the electrons acquire a different effective mass, $m^*_{e,T} = 0.0819 \cdot m_e$\,\cite{Dexter:1956zz}.\\

\section{Electron Drift Model}\label{app:e_drift_model}
In the band structure of germanium, the electrons are populating eight half-ellipsoidal shaped valleys near the edge of the Brillouin zone along the four equivalent $\langle 111\rangle$ directions, see the L-point in Figs.\,\ref{fig:fcc_brillouin}~and~\ref{fig:Ge_Bandstructure}. When viewed in an extended Brillouin scheme\footnote{The  half-ellipsoidal valleys along one axis can be combined.}, four complete ellipsoidal valleys\,\cite{Mihailescu:2000jg,Bruyneel:2006764} remain to be considered. In absence of an electric field, these valleys are equally populated by electrons. Electric fields influence the population probabilities. The electron drift velocity in dependence on the electric field at a given point, $\mathbf{r}$, can be modeled as:
\begin{linenomath}
\begin{equation} \label{eq:eledrifta}
    \mathbf{v}_{e}(\bm{\mathcal{E}}(\bm{r})) = \mathcal{A}(\left|\bm{\mathcal{E}}(\bm{r})\right|)\sum_{j=1}^{4} \frac{n_j}{n} \frac{\gamma_j\bm{\mathcal{E}}_n(\bm{r})}{\sqrt{\bm{\mathcal{E}}_n(\bm{r})^\top\gamma_j\bm{\mathcal{E}}_n(\bm{r})}}~~~~~,
\end{equation}
\end{linenomath}
where $\gamma_j$ is the reciprocal effective mass tensor for the $j$th valley, $\frac{n_j}{n}$ is the fraction of charge carriers in the $j$th valley, $\bm{\mathcal{E}}_n$ is the normalized electric field vector and $\mathcal{A}(\mathcal{E})$ is a function of the magnitude of the electric field.\\
The $\gamma_j$ mass tensors for the respective valleys are obtained by transforming the effective mass tensor, $\gamma_0$,
\begin{linenomath}
\begin{equation}
    \gamma_0 =
    \begin{pmatrix}
        1/m_{e,T}^{*} & 0 & 0 \\
        0 & 1/m_{e,L}^{*} & 0 \\
        0 & 0 & 1/m_{e,T}^{*}
    \end{pmatrix}
\end{equation}
\end{linenomath}
via rotation matrices, $R_j$, from the local coordinates of the valley to crystal coordinates, i.e.\ $\gamma_j=R_{j}^{T}\gamma_0R_{j}$.
When the electric field is aligned with the $\langle100\rangle$ direction, the four valleys are equally populated and $\frac{n_j}{n} = \frac{1}{4}$. For any other direction the relative deviations can be modeled with an additional empirical function\,\cite{Mihailescu:2000jg}:
\begin{linenomath}
\begin{equation}
    \frac{n_j}{n}(\bm{\mathcal{E}}) = \mathcal{R(\mathcal{\bm{E}})} \left( \frac{1/\sqrt{\bm{\mathcal{E}}_n(\bm{r})^\top\gamma_j\bm{\mathcal{E}}_n(\bm{r})}}{\sum\limits_{i=1}^4 1/\sqrt{\bm{\mathcal{E}}_n(\bm{r})^\top\gamma_i\bm{\mathcal{E}}_n(\bm{r})}} - \frac{1}{4}\right) + \frac{1}{4}~~~~~.
\end{equation}
\end{linenomath}
Both $\mathcal{R}(\mathcal{E})$ and $\mathcal{A}(\mathcal{E})$ can be calculated using the velocities along the major axes $\langle100\rangle$ and $\langle111\rangle$, see Eq.\,\eqref{eq:drift_velocity}, and a set of parameters, $\Gamma_{0}$, $\Gamma_{1}$ and $\Gamma_{2}$, which depend only on $m^*_{e,T}$ and $m^*_{e,L}$:
\begin{linenomath}
\begin{align}
    \mathcal{A}(\mathcal{E}) &= \frac{v_e^{100}(\mathcal{E})}{\Gamma_0}~~~~~~~~\,~~~~~~\text{with}~~~ \Gamma_{0} = 2.88847~~~~~,\\
    \mathcal{R}(\mathcal{E}) &= \Gamma_1\frac{v_e^{111}(\mathcal{E})}{\mathcal{A}(\mathcal{E})} + \Gamma_2~~~~~\text{with}~~~ \Gamma_{1} = -1.18211,~~\Gamma_{2} = 3.16066~~~~~.
\end{align}
\end{linenomath}
A detailed derivation of these expressions can be found in Ref.\,\cite{Hagemann19:MT}. 

\section{Response Functions}\label{sec:response_functions}
\begin{figure}[thb]
    \centering
    \includegraphics[width=\textwidth]{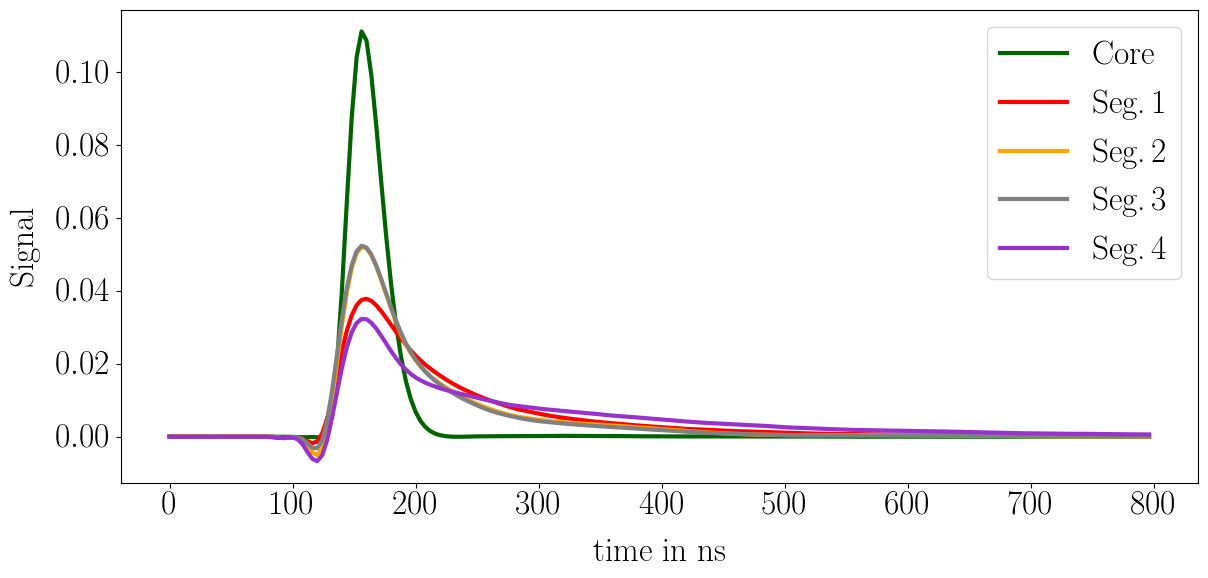}
    \caption[Measured response functions for all the channels for the n-type segmented BEGe.]{Measured response functions for all read-out channels of the segBEGe.}\label{fig:n-type_response_functions}
    \vspace{0.4cm}
\end{figure}

Each preamplifier has an individual so-called response function, i.e.~the response of the system to a Dirac $\delta$-distribution. The shape of the response function is closely related to the bandwidth of the preamplifier. The higher the bandwidth of the preamplifier,
the closer the response function resembles a $\delta$-function
and the less the pulse shape is affected.\\
The preamplifiers were kept at fixed positions in the system and at room temperature in an air-conditioned laboratory for all 
measurements,
such that the response function of each channel was constant.\\
To measure the response function directly, pulses formed like a Dirac-$\delta$ distribution would need to be injected to the test input of the preamplifiers. As this is not feasible, step pulses can be used as input, of which the derivative is a Dirac-$\delta$ distribution. Consequently, the response function is the derivative of the output pulse.\\
Rectangular pulses from a pulse generator were injected into the test inputs of the preamplifiers for each individual channel and the resulting pulses were recorded. The preamplifiers remained installed in the system during the measurement. Thus, the peripheral contributions were accounted for. The rise time of the rectangular pulse was shorter than the sampling time of the ADC and, thus, approximated a step function. Superpulses were formed out of $\approx 20000$ recorded pulses to average out noise and the preamplifier decay was corrected for before taking the derivative.\\
Figure~\ref{fig:n-type_response_functions} shows the resulting response
functions of all channels used for the measurements presented in this paper.
\FloatBarrier
\clearpage
\bibliography{./main}
\end{document}